\documentclass[letterpaper,titlepage,11pt]{article}
\pdfoutput=1
\usepackage{hyperref}

\usepackage{amssymb,amsmath,amsfonts}
\usepackage{graphicx}
\usepackage{xcolor}
\usepackage{epsfig}
\usepackage{array}
\usepackage{booktabs}
\usepackage{subfig}

\usepackage[numbers,merge,sort&compress]{natbib}

\def\clock{{\count0=\time
          \divide\count0 60 
          \ifnum\count0<10 0\fi\the\count0
          \multiply\count0 -60 \advance\count0 \time
          :\ifnum\count0<10 0\fi \the\count0
        }}
\newcommand{\timestamp}{{\small\vbox{\hbox{\tt\jobname.tex}
\hbox{\the\day/\the\month/\the\year, \clock}}}}

\newcommand{\CF}{\mathcal{F}}

\newcommand{\CL}{\mathcal{L}}
\newcommand{\CO}{\mathcal{O}}

\newcommand{\CT}{\mathcal{T}}

\newcommand{\nn}{\nonumber}

\newcommand{\spa}{\, , \qquad }

\newcommand{\beq}{\begin{equation}}
\newcommand{\eeq}{\end{equation}}

\newcommand{\ds}{\displaystyle}

\def\ie{\textit{i.e.} }

\numberwithin{equation}{section}

\begin{document}

\begin{titlepage}
\ \ \vskip 1.8cm
\centerline {\huge \bf The Flowing BIon} 
\vskip 1.4cm

\centerline{\large {\bf Gianluca Grignani$\,^{1}$},  {\bf Troels Harmark$\,^{2}$},}
\vskip 0.2cm \centerline{\large {\bf Callum Kift$\,^{2}$},  {\bf Andrea Marini$\,^{1}$}
and
{\bf Marta Orselli$\,^{1,2}$} }

\vskip 0.7cm

\begin{center}
\sl $^1$ Dipartimento di Fisica e Geologia, Universit\`a di Perugia,\\
I.N.F.N. Sezione di Perugia,\\
Via Pascoli, I-06123 Perugia, Italy
\vskip 0.3cm
\sl $^2$ The Niels Bohr Institute, Copenhagen University  \\
\sl  Blegdamsvej 17, DK-2100 Copenhagen \O , Denmark
\end{center}
\vskip 0.3cm

\centerline{\small\tt grignani@pg.infn.it, harmark@nbi.dk, }
\centerline{\small\tt andrea.marini@pg.infn.it, orselli@nbi.dk}

\vskip 1.2cm \centerline{\bf Abstract} \vskip 0.2cm \noindent
In this paper we use the effective blackfold description of branes to extend the study of the thermal BIon, a D-brane and parallel anti-D-brane connected by a wormhole with F-string charge in hot flat space, by introducing a radial boost along the brane. The boosted system behaves qualitatively differently from both the extremal and the thermal BIon considered previously. 
Interestingly, we are able to formulate a first law of thermodynamics for the system as a whole, despite the fact that it is not a stationary blackfold. In particular, the global temperature is given by the rest frame temperature times the gamma factor of special relativity which is the inverse transformation compared to the case of stationary blackfolds. In addition we define two new kinds of thermodynamic conjugate variables, the energy flux $W$ and the integrated velocity on the brane. We find that a 
phase transition occurs by varying the energy flux $W$. Below a critical value of $W$ the brane separation $\Delta$ changes only slightly with $W$. Instead above the critical value $\Delta$ grows exponentially.

\end{titlepage}

\tableofcontents
\pagestyle{plain}
\setcounter{page}{1}

\section{Introduction}

A BIon is a specific D-brane configuration in which a pair 
of D-brane/anti-D-brane, asymptotically parallel, are connected by a wormhole with F-string charge. 
This configuration was originally discovered as
a solution of the Dirac-Born-Infeld (DBI) action in flat space and at zero
temperature~\citep{Callan:1997kz,Gibbons:1997xz,*Thorlacius:1997zd,*Emparan:1997rt}.

In~\citep{Grignani:2010xm,Grignani:2011mr,Grignani:2012iw,*Armas:2012bk} a novel approach to describe D-branes probing thermal backgrounds was proposed and 
originally applied to the study of the BIon solution for D3/anti-D3-branes in hot flat space.\footnote{See also \cite{Sepehri:2017mqw, *Sepehri:2017gwk,*Sepehri:2016vhy,*Sepehri:2016sjq,*Sepehri:2014jla} for various applications of the static thermal BIon solution of \citep{Grignani:2010xm,Grignani:2011mr}. See \cite{Niarchos:2012pn,Niarchos:2012cy,Niarchos:2013ia} for a M2-M5 generalization of the static thermal BIon solution of \citep{Grignani:2010xm,Grignani:2011mr}. See furthermore \cite{Lunin:2007mj} for a supergravity description of the zero-temperature BIon solution.} The key feature of this new method is that it 
requires thermal equilibrium between the probe and the background and thus it takes into account 
the excitation of the internal degrees of freedom of the brane induced by its surrounding. This is 
achieved by describing the brane probes through the blackfold approach~\citep{Emparan:2009at, Emparan:2007wm,*Emparan:2009cs,*Emparan:2009vd,*Emparan:2009zz,*Emparan:2011hg}.

In this paper we consider a generalization of the hot BIon solution to the case in which there is a velocity along the $r$-direction.\footnote{$r$
denotes the radial direction on the brane world-volume whose origin sits at the center of the wormhole throat.}
In particular we take into account the case of D3/anti-D3-branes, so as to extend the results of \citep{Grignani:2010xm,Grignani:2011mr}.

The extremal, zero temperature, BIon solution is invariant under a boost along the $r$ direction. Locally, one can see this as a consequence of the fact that boosting along the direction of the electric field $\vec{E}$ does not change the electric field.
From the brane point of view it is because boosting along the $r$-direction means that we are boosting along both the F-string and the D3-brane world-volumes, and they are both boost invariant. Hence no new extremal configuration can be found by having a velocity in the $r$-direction. Instead for the non-extremal BIon, we no longer have boost invariance along the F-string and D3-brane world-volumes, hence the configuration with a non-zero velocity is different from the one with zero velocity, but they both have the same extremal limit.

Because of the boosting, the hot BIon system acquires a non-zero energy flow $W$ which has to be radially conserved as a consequence of the blackfold intrinsic equations. The presence of this energy flow is the reason why we refer to this new configuration as the ``flowing BIon''. 

An important motivation for this study is that our flowing BIon solution gives a new example of a black brane configuration that has an event horizon that is not a Killing horizon. The reason that it cannot be a Killing horizon is that the blackfold then should be a stationary blackfold with a world-volume Killing vector field (KVF) along the fluid velocity \cite{Caldarelli:2008mv,Emparan:2009at}. 
But such a KVF field along the fluid velocity means that a heat flow is impossible. Therefore, we demonstrate that one can use the blackfold effective description of black branes to find new brane configurations with heat flow that does not possess a Killing horizons. This means that one can use the blackfold formalism to find new types of black brane configurations in higher dimensions of this kind.

Another motivation for this study is that solutions with heat flow arise in the context of the AdS/CFT correspondence. 
For instance in Refs \cite{Fischetti:2012vt}, \cite{Fischetti:2012ps} and \cite{Hubeny:2009kz}
a boundary spacetime which contains a pair of black holes connected through the bulk by a tubular bulk horizon was considered. 
Taking one boundary black hole to be hotter than the other prohibits equilibrium. The result is a so-called ``flowing funnel'',
a stationary bulk black hole with a non-Killing horizon that may be said to transport heat toward the cooler boundary black hole. 
Furthermore, within the context of black holes in asymptotically flat space an explicit example of a ``flowing black funnel" was constructed in any dimension $D\ge 5$ in \cite{Emparan:2013fha}.

In this paper we consider the leading order blackfold approach to black brane configurations with heat flow and apply it to find the solution for the flowing BIon. Thus, we do not include the higher-order effects of dissipation which is a key feature of non-stationary fluids. This means that our flowing BIon configuration is actually stationary, despite the fact that it does not possess a Killing horizon. For this reason the heat reservoirs that the heat is transported from and to are at the same temperature. It would be highly interesting to include the effect of dissipation and see how this effects our flowing BIon configuration.

The blackfold analysis we perform for this system allows us to find a ``new global thermodynamics'' 
for a solution with heat flow, \ie the thermodynamics that should correspond to a free energy 
for a system with velocity $V$ and heat flow $W$. 
In the $W=0$ case the free energy $\CF$ satisfies the first law of thermodynamics $d \CF = - S dT$, 
where $S$ is the entropy and $T$ the global temperature.  
To find the thermodynamics for non-zero $W$ we should add a term corresponding to the variation of $W$, 
\ie  $d\CF = - S dT - V dW$ where $V$ is the conjugate variable to $W$. 
In the BIon system we consider, $W$ is the total energy flux passing through a given surface of constant $r$. 
$W$ is an extensive variable with respect to the two-sphere directions (transverse to the F-string) 
and an intensive variable with respect to the radial direction $r$ (parallel to the F-string). 
The conjugate variable $V$ should thus be intensive with respect to the two-sphere directions and 
extensive with respect to the radial direction; we will show that $V$ turns out to be the integrated velocity 
along the brane world-volume.

The global temperature $T$ can be interpreted as a relativistic transformation of the local temperature 
in the comoving frame (\ie  the rest frame), that we denote as $\CT$, to the frame in which one observes a velocity. 
We find that the relation between $T$ and $\CT$ is given by $T =  \CT/\sqrt{1-v^2}$. 
Note that this transforms oppositely compared with what we are used to for stationary blackfolds. 
However, for stationary blackfolds there is no actual heat transport since one has a symmetry along the heat flow. 
Instead, for our stationary system one needs two heat reservoirs from which to extract and transfer heat. 
Thus, it is a highly different type of system, as seen thermodynamically.

It is interesting to note that in the literature there are several schools of thought on how the temperature should transform from the rest frame to a moving frame in the theory of relativity. The classical view by Einstein, Planck, Pauli and many other physicists, starting from 1910's, is that $T_{\rm moving} = \sqrt{1-v^2} T_{\rm rest}$ where $T_{\rm rest}$ is the rest-frame temperature while $T_{\rm moving}$ is the temperature in a moving frame.  A more modern view, starting from the 1950's, is that $T_{\rm moving} =  T_{\rm rest}/\sqrt{1-v^2}$, as first advocated by Ott, Arzeli\`es and M\o ller. Einstein himself later adopted this view as well \cite{Przanowski:2010,Callen:1971}.  Amusingly, we see that the transformation law of the stationary blackfold corresponds to the ``classical" transformation $T_{\rm moving} = \sqrt{1-v^2} T_{\rm rest}$ while the transformation for a blackfold with heat flow as considered in this paper gives $T_{\rm moving} =  T_{\rm rest}/\sqrt{1-v^2}$.

The paper is organized as follows.
In Section \ref{sec:fluiddynamics} we consider the fluid description of the black F1-D3-brane bound state solution,
which is the supergravity solution we use to describe the BIon system through the blackfold approach.
Then, in Section \ref{sec:setup} we present the ansatz we consider for the embedding of the F1-D3 system in hot
flat space.

Section \ref{sec:blackfoldeqs} is devoted to the analysis of the blackfold equations for the flowing BIon configuration.
We show that the intrinsic blackfold equations, which are given by the energy-momentum conservation, 
imply that the energy flux $W$ and the global temperature $T$ are constant along the brane wold-volume.
The extrinsic equations instead govern the behavior of the embedding functions of the F1-D3-brane bound state and so
fix the actual shape of the BIon system. 
From the analysis of the blackfold equations emerges that the system undergoes a phase transition at a certain value of the 
energy flux, $W_{\rm crit}$. 
It turns out that when the energy flux is below a critical value, $W<W_{\rm crit}$, the behavior of the system is 
qualitatively identical to the non-flowing solution of \cite{Grignani:2010xm}. 
For  $W>W_{\rm crit}$ instead we observe a severe change with respect to the $W=0$ case.

In Section \ref{sec:action} we derive an action principle for the extrinsic equations governing the flowing BIon system
by generalizing the free energy of the hot BIon solution with $W=0$ to non-zero $W$.
This also allows us to study the thermodynamics in the presence of heat flow.
Finally we use the free energy as thermodynamic potential to study the stability of the solutions.

Section \ref{sec:conclusion} contains the conclusions and in Appendix \ref{app:fluidtemp} we discuss further on the relativistic fluid temperature 
in the presence of a non-zero fluid velocity by taking into account a simpler setup.

\section{Fluid description of F1-D3-brane bound state}
\label{sec:fluiddynamics}

We begin by considering the fluid description of the F1-D3-brane bound state. This will be necessary later for formulating the intrinsic and extrinsic blackfold equations that we shall solve for the case of a BIon system of branes with a heat flow.

\subsection{Local quantities and thermodynamics}

Consider $N$ D3-branes with an electric field corresponding to $k$ units of electric flux at non-zero temperature. The metric for such a system is that of a black D3-F1 brane bound state, which in the regime of large $N$ and in the string frame, reads \cite{Harmark:2000wv}
\begin{equation}
\label{D3F1_geom}
ds^2 = D^{-\frac{1}{2}} H^{-\frac{1}{2}} ( - f dt^2 + dx_1^2 ) +
D^{\frac{1}{2}} H^{-\frac{1}{2}} ( dx_2^2 + dx_3^2 ) +
D^{-\frac{1}{2}} H^{\frac{1}{2}} ( f^{-1} dr^2 + r^2 d\Omega_5^2 ) \, ,
\end{equation}
where\footnote{From here on $\alpha=\alpha(r)$, $\zeta=\zeta(r)$ and $r_0=r_0(r)$.}
\begin{equation}
\label{fHD}
f = 1 - \frac{r_0^4(r)}{r^4} \spa H = 1 + \frac{r_0^4(r) \sinh^2
\alpha(r)}{r^4} \spa D^{-1} = \cos^2 \zeta(r) + \sin^2 \zeta(r) H^{-1}\, ,
\end{equation}
and with dilaton field $\phi$, Kalb-Ramond field $B_{(2)}$, and two- and four-form Ramond-Ramond gauge fields $C_{(2)}$ and $C_{(4)}$ given by
\begin{equation}
\label{D3F1fields}
\begin{array}{c} \ds
e^{2\phi} = D^{-1} \spa B_{01} = \sin \zeta ( H^{-1} -1 ) \coth \alpha \\[4mm] \ds
C_{23} = \tan
\zeta ( H^{-1} D  - 1) \spa C_{0123} = \cos \zeta D ( H^{-1} - 1 ) \coth \alpha \, .
\end{array}
\end{equation}

We now want to derive the energy-momentum tensor for this system with a non-trivial world-volume metric. We follow the analysis of \cite{Caldarelli:2010xz,Emparan:2011hg}.
Denoting the world-volume coordinates as $\sigma^a$ and the world-volume metric  
as $\gamma_{ab}$ ($a,b=0,1,2,3$), we can write the general form of the energy-momentum tensor for the black F1-D3 brane bound state as \cite{Grignani:2010xm,Caldarelli:2010xz,Emparan:2011hg}
\begin{equation}\label{Tab}
T^{ab} = \epsilon u^a u^b + P_\parallel v^a v^b + P_\perp \Delta_{ab} \spa \Delta_{ab} = \gamma^{ab} + u^a u^b - v^a v^b\, .
\end{equation}

The strings in the BIon setup break the full isotropy and consequently in \eqref{Tab} we introduced the parallel ($P_\parallel$) and orthogonal ($P_\perp$) pressure. Moreover we require the vector fields $u^a$ and $v^a$ to be orthonormal
\begin{equation}
\label{usqr}
\gamma^{ab} u_a u_b = -1 \spa \gamma^{ab} v_a v_b = 1 \spa \gamma^{ab} u_a v_b = 0\, .
\end{equation}

Reading off the energy-momentum tensor as seen by an asymptotic observer from the metric \eqref{D3F1_geom} (see for instance  ref.~\citep{Harmark:2004ch})
we find the expression for $\epsilon$, $P_\parallel$ and $P_\perp$ in \eqref{Tab}
\begin{equation}
\begin{split}
\epsilon &= \frac{\pi^2}{2} T_{\rm D3}^2 r_0^4 (5+4\sinh^2 \alpha) \spa P_\parallel = -\frac{\pi^2}{2} T_{\rm D3}^2 r_0^4 (1+4\sinh^2 \alpha)\, ,  \\
P_\perp &= -\frac{\pi^2}{2} T_{\rm D3}^2 r_0^4 (1+4 \cos^2 \zeta \sinh^2 \alpha)\, .
\end{split}
\end{equation}

The F-string and D3-brane currents are
\begin{equation}
J^{\rm (F1)} = q_1 \, u \wedge v \, , \qquad
J^{\rm (D3)} = q_3\, \sqrt{\gamma} \, d\sigma^0 \wedge d\sigma^1  \wedge d\sigma^2  \wedge d\sigma^3 \, ,
\end{equation}
where $\sqrt{\gamma}$ is the determinant of the world-volume metric and $q_1$ and $q_3$ are the associated charge densities\footnote{We used here that $4\Omega_5 / (16\pi G) = 2\pi^2 T_{\rm D3}^2$.}
\begin{equation}
q_1 = 2 \pi^2 T_{\rm D3}^2 r_0^4 \sin \zeta \cosh \alpha \sinh \alpha\, , \qquad q_3 =  2 \pi^2 T_{\rm D3}^2 r_0^4 \cos \zeta \cosh \alpha \sinh \alpha \, .
\end{equation}
The conjugated chemical potentials are
\begin{equation}
\mu_1 = \sin \zeta \tanh \alpha\, , \qquad  \mu_3 = \cos \zeta \tanh \alpha\, .
\end{equation}

The local temperature $\CT$ (the temperature measured by a co-moving observer, \ie in a rest frame) is
\begin{equation}
\label{localtemp}
\CT = \frac{1}{\pi r_0 \cosh \alpha}\, .
\end{equation}
The entropy density is
\begin{equation}
s = 2\pi^3 T_{\rm D3}^2 r_0^5 \cosh \alpha \, .
\end{equation}

For small temperatures $\CT$ (near extremality) we find
\begin{equation}
\begin{split}
\epsilon &= \frac{N T_{\rm D3}}{\cos \zeta} + \frac{3\pi^2 N^2 \CT^4}{8 \cos ^2\zeta} + \CO(\CT^8)
\spa
P_\parallel = - \frac{N T_{\rm D3}}{\cos \zeta} + \frac{\pi^2 N^2 \CT^4}{8 \cos ^2\zeta} + \CO(\CT^8) \, ,\\
P_\perp &= - N T_{\rm D3}\cos \zeta + \frac{\pi^2 N^2 \CT^4}{8} \Big( 2 - \frac{1}{\cos^2 \zeta} \Big) + \CO(\CT^8) \, .
\end{split}
\end{equation}

There are several relations that the fluid should obey, most of which are easily verified; one of these is the
Gibbs-Duhem relation
\begin{equation}
\label{gibbsduhem}
\epsilon + P_\parallel = \CT s\, .
\end{equation}

It is straightforward to check that the pressure satisfy the following relation
\begin{equation}
\label{pressurerelation}
P_\perp = P_\parallel + \mu_1 q_1\, ,
\end{equation}
which tells us that the pressure difference is the result of the energy density of the strings, $\mu_1 q_1$.

The local first law of thermodynamics reads
\begin{equation}
\label{localfirstlaw}
d\epsilon = \CT ds + \mu_1 dq_1 + \mu_3 d q_3 \, .
\end{equation}
Combining this with \eqref{gibbsduhem} and \eqref{pressurerelation} we get
\begin{equation}
\label{firstlawpressures}
dP_\parallel = s d\CT - \mu_1 dq_1  - \mu_3 d q_3 \spa
dP_\perp =  s d\CT + q_1 d\mu_1  - \mu_3 d q_3  \, .
\end{equation}

\subsection{Conservation laws}

The charge currents introduced in the previous subsection obey the conservation laws 
\begin{equation}
\label{chargecurrentcons0}
D_a J_{\rm F1}^{ab} = 0 \spa D_a J_{\rm D3}^{abcd}  = 0 \, ,
\end{equation}
where $D_a$ is the world-volume covariant derivative.
We can write these as
\begin{equation}
\label{chargecurrentcons}
\partial_a( \sqrt{\gamma} J_{\rm F1}^{ab} ) = 0 \spa \partial_a ( \sqrt{\gamma} J_{\rm D3}^{abcd} ) = 0 \, .
\end{equation}
Since $J_{\rm D3}^{abcd} = - \epsilon^{abcd} q_3 /\sqrt{\gamma}$
the D3-brane current conservation law can thus be written
\begin{equation}
\label{D3cons}
\partial_a q_3 = 0 \, .
\end{equation}
This is consistent with the quantization of the D3-brane charge $q_3 = N T_{\rm D3}$, \ie
\begin{equation}
\label{Nquant}
N = 2\pi^2 T_{\rm D3} r_0^4 \cos \zeta  \cosh \alpha \sinh \alpha \, ,
\end{equation}
since this means $\partial_a N = 0$.
Projecting the F-string current conservation law onto the $u^a$ and $v^a$ directions, respectively, it is equivalent to the two equations%
\footnote{To see this use that $J_{\rm (F1)}^{ab} = q_1 (u\wedge v)^{ab} = q_1 ( u^a v^b - u^b v^a)$ and compute $u_b D_a J_{\rm (F1)}^{ab}$ and $v_b D_a J_{\rm (F1)}^{ab}$, both of which should be zero by \eqref{chargecurrentcons0}.}
\begin{equation}
\label{F1cons}
D_a (q_1 v^a ) = q_1 v^b u^a D_a u_b \spa D_a ( q_1 u^a ) = - q_1 u^b v^a D_a v_b \, .
\end{equation}

The energy-momentum conservation law, which corresponds to the intrinsic blackfold equations, is
\begin{equation}
\label{EMcons}
D_a T^{ab} =0\, .
\end{equation}
In the following we project this equation onto $u_b$. To this end, we record
\begin{equation}
u_b T^{ab} =  - \epsilon u^a
\spa
T_{ab} D_a u_b =  \mu_1 q_1 u_b v^a D_a v^b + P_\perp D_a u^a\, .
\end{equation}
Hence we find
\begin{equation}
u_b D_a T^{ab} = - D_a ( \epsilon u^a) -  \mu_1 q_1 u_b v^a D_a v^b - P_\perp D_a u^a\, . 
\end{equation}
Using \eqref{gibbsduhem} and \eqref{F1cons} we find
\begin{eqnarray}
&& u_b D_a T^{ab} = - D_a ( \CT s u^a) + D_a ( P_\parallel u^a ) +  \mu_1 D_a ( q_1 u^a) - P_\perp D_a u^a\, , \nn \\ &&= - \CT D_a (s u^a) + u^a [ - s \partial_a \CT + \partial_a P_\parallel + \mu_1 \partial_a q_1 ] + [ P_\parallel + \mu_1 q_1 - P_\perp ] D_a u^a\, .
\end{eqnarray}
Using \eqref{pressurerelation} and \eqref{firstlawpressures} we get
\begin{equation}
u_b D_a T^{ab} = - \CT D_a (s u^a) - u^a \mu_3 \partial_a q_3\, .
\end{equation}
Finally, using \eqref{D3cons}, we get
\begin{equation}
u_b D_a T^{ab} = - \CT D_a (s u^a)\, .
\end{equation}
Thus, as a consequence of the conservation of energy-momentum \eqref{EMcons} we find that the entropy current $s u^a$ is conserved
\begin{equation}
\label{entropycons}
D_a ( s u^a ) = 0\, .
\end{equation}
This generalizes the conservation of entropy current considered for blackfolds with charges in \cite{Caldarelli:2010xz,Emparan:2011hg} to include both an F1-string and a D3-brane charge.

\section{Setup for BIon configuration}
\label{sec:setup}

In this section we introduce the setup that describes the BIon solution. We define what type of embedding we consider for the D3-brane, what are the boundary conditions that we impose and what conserved charges we have.

\subsubsection*{Metric and boundary conditions}

We embed the D3-brane world volume in 10D Minkowski space-time with metric
\begin{equation}
ds^2 = -dt^2 + dr^2 +r^2 (d\theta^2 + \sin^2 \theta d\phi^2) + dz^2 + \sum_{i=1}^5 dx_i^2\, .
\end{equation} 
The ansatz we choose for the embedding is 
\begin{equation}
t(\sigma) = \sigma^0 \spa r(\sigma) = \sigma^1 \spa z(\sigma) = z(\sigma^1) \spa \theta(\sigma)=\sigma^2 \spa \phi(\sigma)=\sigma^3\, ,
\end{equation}
while the remaining coordinates, \ie $x_{i=2,3,\dots,6}$, are constant. The function $z(\sigma)$ describes the bending of the brane. The induced metric on the brane then is
\begin{equation}
\label{indmet1}
\gamma_{ab} d\sigma^a d\sigma^b = - dt^2 + (1+ {z'}^2) dr^2 + r^2 (d\theta^2 + \sin^2 \theta d\phi^2 )\, ,
\end{equation}
where $z=z(r)$.
In order to have a BIon configuration we have to impose the following boundary conditions
\begin{equation}\label{bc}
z(r)\to 0~~\text{for}~~r\to\infty,~~~~z'(r)\to-\infty~~\text{for}~~r\to r_{\rm min}\, ,
\end{equation}
which means that the brane is asymptotically flat and the throat has a minimal radius $r_{\rm min}$. 

The defining feature of our configuration is the local velocity $v(r)$. This is the velocity tangential to the brane surface and radially directed along the string that is dissolved into the brane.
To define $v(r)$ we should give it an invariant geometric meaning, \ie we should measure it with respect to a local Lorentz frame. Since $\gamma_{tt}=-1$ and the metric is diagonal we know that this means that $u_0$ is just the Lorentz factor, $u_0 = 1/ \sqrt{1-v^2}$ hence we can use this as the defining relation for $v$. Solving \eqref{usqr} we find%
\begin{equation}
\begin{array}{c} \ds
u_0 = \frac{1}{\sqrt{1-v^2}} \spa u_1 = \frac{v \sqrt{1+{z'}^2}}{\sqrt{1-v^2}} \spa u_2=u_3=0
\\[5mm] \ds
v_0 = \frac{v}{\sqrt{1-v^2}} \spa v_1 = \frac{\sqrt{1+{z'}^2}}{\sqrt{1-v^2}} \spa v_2 = v_3 = 0
\end{array}
\end{equation}
where $v=v(r)$. This gives \eqref{usqr} since $\gamma^{11} = 1/(1+{z'}^2)$.

\subsubsection*{Conservation of the F-string and D3-brane charge currents}

With the setup described above, the conservation of the D3-brane charge current \eqref{D3cons} takes the form $\partial_r q_3 = 0$, or, equivalently, $\partial_r N =0$.

Note that
\begin{equation}
( u \wedge v)_{01} = u_0 v_1 - v_0 u_1 = \sqrt{1+{z'}^2}\, .
\end{equation}
Hence 
\begin{equation}
u \wedge v \wedge r^2 \omega_2 = \sqrt{\gamma} \, d\sigma^0 \wedge d\sigma^1 \wedge d\sigma^2 \wedge d\sigma^3\, ,
\end{equation}
where $\omega_2 = \sin \theta d\theta \wedge d\phi$ is the volume element of the unit two-sphere.

Consider the conservation law for the F-string current \eqref{chargecurrentcons}.
Since $r_0$, $\zeta$ and $\alpha$ only depends on $r$ this gives
\begin{equation}
\partial_r ( \sqrt{\gamma} \sqrt{\gamma^{11}} q_1 )  = 0\, .
\end{equation}
Using $\partial_r q_3=0$ we see this is equivalent to
\begin{equation}
\label{consF1}
\partial_r ( r^2 \tan \zeta )  = 0\, .
\end{equation}
Hence it is consistent with charge quantization for the F-string
\begin{equation}
\label{kF1}
k T_{\rm F1} = N T_{\rm D3} \int d\sigma^2 d\sigma^3 \sqrt{\gamma_{22}\gamma_{33}} \tan \zeta = 4\pi r^2 N T_{\rm D3} \tan \zeta\, .
\end{equation}
This gives
\begin{equation}
\label{kappa_zeta}
\cos \zeta = \frac{1}{\sqrt{1+\frac{\kappa^2}{r^4} }} \spa \kappa \equiv \frac{k T_{\rm F1}}{4\pi N T_{\rm D3}}\, ,
\end{equation}
where $k$ is some integer and corresponds to the number of F-strings

\subsubsection*{Comments on stationarity of BIon solution}

Our BIon solution is stationary in the sense that all fields are independent of the time $t$.
Specifically, $r_0$, $\alpha$ and $\zeta$ are independent of time, consequently there is a Killing vector field (KVF) $\partial/\partial t$.
However, it is clearly not static, as it does transport heat from one place to another. Namely, it transports heat from $r=\infty$ for brane 1 towards $r_{\rm min}$, and again to $r=\infty$ for brane 2 (we assume the separation of the two branes is finite). This can be a stationary configuration since we also assume that there are two heat reservoirs (AKA thermal reservoirs), both at temperature $T$, at $r=\infty$ for brane 1 and at $r=\infty$ for brane 2. Per definition a heat reservoir has a very large heat capacity such that at the time scale for which we are considering the solution the heat reservoir can stay at a fixed temperature no matter how much heat we draw from it.
The other reason why it can be a stationary solution is that, as mentioned above, we are working in the leading order blackford formalism that does not take into account dissipation. It would be interesting to see how the higher-order effect of dissipation can alter the system.

\section{The blackfold equations}
\label{sec:blackfoldeqs}
	
In this section we shall solve the equations governing the blackfold dynamics.
These can be divided in two sets: 
The first one contains the so called intrinsic equations, \eqref{EMcons}, which coincide with the energy momentum conservation. 
We shall see that for our setup they give rise to two independent equations that physically state the radial conservation of 
the energy flux and of the global temperature.
The other set of blackfold equations is formed by the extrinsic equations, $T^{ab} K_{ab} {}^\mu = 0$, 
which for our system also reduce to only two equations since the $b=\theta,\phi$ components are
trivially satisfied. 
We start by focusing on the intrinsic equations.

\subsection{The intrinsic equations}

For the induced metric \eqref{indmet1} we have 
the non-zero components of the Christoffel symbols
\begin{equation}
\begin{array}{c}\ds
\Gamma^{1}_{11} = \frac{1}{2} \partial_r \log \gamma_{11} \ , \  
\Gamma^a_{a1} =  \frac{1}{2} \partial_r \log \gamma_{11}+ \frac{2}{r}\ , \  \Gamma^{1}_{22} = - r \gamma^{11} \ , \  \Gamma^{1}_{33} = - r \sin^2\theta \gamma^{11} \, ,
\\[3mm] \ds
 \Gamma^2_{2 1} = \Gamma^3_{3 1} =\frac{1}{r} \ , \  
 \Gamma^2_{33} = - \sin \theta \cos \theta
\ , \ 
 \Gamma^3_{3 2}  = \Gamma^a_{a2} =  \cot \theta
\ , \ 
\Gamma^a_{ab} = \partial_b \log \sqrt{\gamma}\, .
\end{array}
\end{equation}
We can write \eqref{EMcons} as
\begin{equation}
0 = D_a T^{ab} = \partial_a T^{ab} + \Gamma^a_{ac} T^{cb} + \Gamma^b_{ac} T^{ac} = \frac{1}{\sqrt{\gamma}} \partial_a ( \sqrt{\gamma} T^{ab} )  + \Gamma^b_{ac} T^{ac}\, .
\end{equation}
The time component of \eqref{EMcons} is
\begin{equation}
\label{EMconstime}
0 = D_a T^{a0} = \frac{1}{\sqrt{\gamma}} \partial_r ( \sqrt{\gamma} T^{10} ) \, ,
\end{equation}
and for the $r=\sigma^1$ component we find
\begin{equation}
0 = D_a T^{a1} = \frac{1}{\sqrt{\gamma}}  \partial_r (\sqrt{\gamma} T^{11})  + T^{11} \partial_r \log \sqrt{\gamma_{11}} - \frac{2}{r} \gamma^{11} P_\perp = \frac{1}{r^2} \gamma^{rr} \Big[ \partial_r ( r^2 \gamma_{11} T^{11} ) - 2r P_\perp \Big]\, .
\end{equation}
Hence this conservation equation can be written
\begin{equation}
\label{EMconsr}
\partial_r ( r^2 \gamma_{11} T^{11} ) = 2r P_\perp\, .
\end{equation}
The angular components $\theta=\sigma^2$ and $\phi=\sigma^3$ are
\begin{equation}
D_a T^{a2} = \Gamma^a_{a2} T^{2 2} + \Gamma^2_{ac} T^{ac} = \cot \theta P_\perp \frac{1}{r^2} + (- \sin \theta \cos \theta) P_\perp \frac{1}{r^2 \sin^2\theta} = 0\, ,
\end{equation}
\begin{equation}
D_a T^{a3} = \Gamma^a_{a3} T^{2 2} + \Gamma^3_{ac} T^{ac} = 0\, ,
\end{equation}
which therefore are both trivially solved.

\subsubsection*{Conservation of energy flux}

We saw that the time component of \eqref{EMcons} gives \eqref{EMconstime}. As we now shall see, this conservation equation means
that the energy flux through each surface of constant $r$ is conserved along $r$. 

The momentum density is $T^{j0}$. Instead the energy flux is $T^{0j}$. These two are equal since the theory is relativistic.%
\footnote{This is because $E=mc^2$ thus the momentum associated with particles moving can be mapped directly to the transfer of energy (see \cite{misner1973gravitation}). 
	}
We choose here to interpret it as the energy flux, \ie
\begin{equation}
n_a T^{0a} dA \, ,
\end{equation}
is the energy per unit time passing through an area $dA$ perpendicular to the unit normal vector $n^a$. We are interested in the energy passing through the surfaces of constant $r$ towards the center of the BIon, hence the unit normal is $n_a = \delta_{a,1} n_1$ with $n_1^2 \gamma^{11} = 1$ and hence $n_1 = - \sqrt{\gamma_{11}}$. To find the total flux passing through the surface of constant $r$s we integrate $\int d\theta d\phi \, r^2 \sin \theta \, n_1 T^{10} = 4\pi r^2 \sqrt{1+{z'}^2}  T^{10}$. Thus the total energy flux for a surface of constant $r$ is
\begin{equation}
\label{Wdef}
W = 4\pi r^2 n_1 T^{10} = \frac{v}{1 - v^2} 4\pi r^2  ( \epsilon+P_{\parallel} )= \frac{v}{1 - v^2} 4\pi r^2  \CT s = 8\pi^3 T_{\rm D3}^2 \frac{v r^2  r_0^4}{1-v^2}\, .
\end{equation}
We see that the conservation equation \eqref{EMconstime} precisely gives
\begin{equation}
\label{Wcons}
\partial_r W = 0\, ,
\end{equation}
\ie that the energy flux is conserved throughout the BIon solution.

\subsubsection*{Conservation of entropy current and global temperature}

For the BIon solution we are considering, the local velocity of the fluid $u^a$ is not proportional to a KVF as it should be for a stationary blackfold. Therefore, it is not a priori clear that one can define a global temperature for the system. We show below that despite the fact that our BIon system is not a stationary blackfold one can nevertheless define a global temperature that is conserved everywhere on the BIon.

Combining the time and radial components, \eqref{EMconstime} and \eqref{EMconsr}, of the intrinsic equations one gets the conservation of the entropy current \eqref{entropycons}. This takes the form
\begin{equation}
\partial_r ( \sqrt{\gamma} s u^r ) = 0\, .
\end{equation}
This is equivalent to 
\begin{equation}
\label{scons}
\partial_r \Big(  \frac{r^2 v s}{\sqrt{1-v^2}}  \Big) =0\, .
\end{equation}
Note that, as pointed out before, the conservation law $D_a T^{ab}=0$ only has two non-trivial components which can be expressed as \eqref{Wcons} and \eqref{scons} (since these are independent of each other). We can combine \eqref{Wcons} and \eqref{scons} as follows
\begin{equation}
0 = \partial_r W - \frac{\CT}{\sqrt{1-v^2}} \partial_r \Big(  \frac{4\pi r^2 v s}{\sqrt{1-v^2}}  \Big) =\frac{4\pi r^2 v s}{\sqrt{1-v^2}} \partial_r \Big(  \frac{\CT}{\sqrt{1-v^2}} \Big)\, 
\end{equation}
This means that instead of imposing entropy current conservation  \eqref{scons} we can equivalently impose 
\begin{equation}
 \partial_r \Big(  \frac{\CT}{\sqrt{1-v^2}} \Big) = 0\, .
 \end{equation}
Thus, we see that we can define what we call the global temperature $T$ as
\begin{equation}
\label{globT}
T = \frac{\CT}{\sqrt{1-v^2}}\, .
\end{equation}
For the BIon configuration the energy-momentum conservation \eqref{EMcons} can then simply be expressed as
\begin{equation}
\partial_r W = 0\, , \qquad \partial_r T = 0\, ,
\end{equation}
\ie the energy flux $W$ and the global temperature $T$ are constant throughout the solution.


The global temperature can be interpreted as a relativistic transformation of the temperature $\CT$ in the comoving frame (\ie  the rest frame) to the frame in which one observes the velocity $u^a$. Note that this transforms oppositely compared with what we are used to for what we call stationary blackfolds \cite{Emparan:2009at}. However, stationary blackfolds means that we have a KVF for the fluid on the brane (which the fluid velocity is proportional to). For such fluids there is no actual heat transport since one has a symmetry along the heat flow. Instead, for our stationary system one needs two heat reservoirs for which to extract and transfer heat. Thus, it is a highly different type of system, as seen thermodynamically. See also Appendix \ref{app:fluidtemp} where this is considered in a simpler setting.

\subsection{The extrinsic equations}

Along with the intrinsic equations considered before, the BIon configuration should satisfy also 
another set of blackfold equations, namely the extrinsic equations, which read  
\begin{equation}
T^{ab} K_{ab} {}^\mu = 0\spa K_{ab} {}^\mu=D_aD_b X^\mu+\Gamma^\mu_{\nu\rho} D_a X^\nu D_b X^\rho\, ,
\end{equation}
where $K_{ab} {}^\mu$ is the extrinsic curvature tensor.
The non vanishing components of  $K_{ab} {}^\mu$ are
\begin{equation}\label{extrinsiccurv}
K_{11} {}^z = \frac{z''}{1+{z'}^2} \spa \gamma^{22} K_{22}{}^z = \gamma^{33} K_{33}{}^z = \frac{z'}{r(1+{z'}^2)} \spa K_{ab}{}^r = - z' K_{ab}{}^z\, .
\end{equation}
From the last equation in \eqref{extrinsiccurv}  we see that it is enough to consider $T^{ab} K_{ab} {}^z = 0$.
Thus we have in practice only one relevant extrinsic equation, which is given by
\begin{equation}
0= \frac{z''}{1+{z'}^2} T^{11} + P_\perp ( \gamma^{22} K_{22}{}^z + \gamma^{33} K_{33}{}^z) = \frac{z''}{1+{z'}^2} T^{11} + 2P_\perp \frac{z'}{r(1+{z'}^2)}\, .
\end{equation}
Using the conservation equation \eqref{EMconsr} in the $r=\sigma^1$ direction this can be written as
\begin{equation}
\frac{z''}{1+{z'}^2} T^{11} +\partial_r ( r^2 \gamma_{11} T^{11} )\frac{z'}{r^2(1+{z'}^2)} = 0\, ,
\end{equation}
which is equivalent to
\begin{equation}
\label{extr35}
\frac{z''}{(1+{z'}^2)^{3/2}} r^2\gamma_{11}T^{11} +\partial_r ( r^2 \gamma_{11} T^{11} )\frac{z'}{\sqrt{1+{z'}^2}} = 0\, .
\end{equation}
Using
\begin{equation}
\left( \frac{z'}{\sqrt{1+{z'}^2}} \right)' = \frac{z''}{(1+{z'}^2)^{3/2}}
\end{equation}
we see that $T^{ab} K_{ab} {}^z = 0$ becomes
\begin{equation}
\label{extreq}
\left( \frac{z'}{\sqrt{1+{z'}^2}}r^2\gamma_{11}T^{11}  \right)' = 0\, .
\end{equation}
This is the extrinsic equation for the BIon that we shall solve below.

Defining
\begin{equation}
\label{Ffct}
F(r) = \frac{4\pi}{W} r^2\gamma_{11}T^{11} = \frac{4\pi r^2}{W} \Big( P_\parallel + \frac{v^2}{1-v^2} \CT s \Big) = \frac{3+v^2}{4v} - \frac{1-v^2}{v} \cosh^2 \alpha
\end{equation}
and keeping into account  that the energy flux is conserved, \eqref{Wcons}, we can write \eqref{extreq} as
 \begin{equation}
\label{extreq2}
\left( \frac{z' F(r)}{\sqrt{1+{z'}^2}} \right)' = 0\, ,
\end{equation}
where we multiplied with the constant $4\pi /W$ to simplify $F(r)$.
Imposing the boundary conditions \eqref{bc} we can solve \eqref{extreq2} as
\begin{equation}
\label{zofr}
z(r) = \int_{r}^\infty dr' \left( \frac{F(r')^2}{F(r_{\rm min})^2} -1 \right)^{-\frac{1}{2}}\, .
\end{equation}
We shall now compute $F(r)$. We show in the following that it can be computed when given the following quantities $W$, $T$, $k$ and $N$. In particular this entails that $F(r)$ does not depend on $z(r)$ and hence \eqref{zofr} can provide the BIon configuration when  $W$, $T$, $k$, $N$ and $r_{\rm min}$ are given.

Use first \eqref{localtemp}, \eqref{Nquant} and \eqref{globT} to get
\begin{equation}
\label{alphaeq}
\frac{\pi^2 N T^4}{2 T_{\rm D3}}  = \frac{\cos \zeta}{(1-v^2)^2}  \frac{\sinh \alpha}{\cosh^3 \alpha}\, .
\end{equation}
We will use this equation to obtain $\alpha$ in terms of $T$, $\zeta$ and $v$. Note first that $\sinh \alpha / \cosh^3\alpha$ is bounded from above by $2\sqrt{3}/9$. Define therefore the angle $\delta(r)$ as
\begin{equation}
\cos \delta  = \frac{3\sqrt{3}}{2}  \frac{\sinh \alpha}{\cosh^3 \alpha}\, ,
\end{equation}
with $0 \leq \delta(r) \leq \pi/2$. We can invert this relation and obtain three possible branches of solutions. 
A similar analysis was performed in \cite{Grignani:2010xm,Grignani:2011mr} for the static configuration. Also in this case one finds that
the solution-branch that is connected to the extremal F1-D3 brane bound state is
\begin{equation}
\label{alphasol}
\cosh^2 \alpha = \frac{3}{2} \frac{\cos \frac{\delta}{3} +\sqrt{3} \sin \frac{\delta}{3}}{\cos \delta} = 3 \frac{\sin ( \frac{\delta}{3}+\frac{\pi}{6} )}{\cos \delta}\, .
\end{equation}
As $\delta$ goes from $0$ to $\pi/2$, $\cosh^2\alpha$ increases monotonically from $3/2$ to $\infty$. We can now write
\eqref{alphaeq} as
\begin{equation}
\label{T_delta}
\overline{T}^4 =  \frac{\cos \zeta}{(1-v^2)^2} \cos \delta\, ,
\end{equation}
where we defined the rescaled temperature 
\begin{equation}
\overline{T} = \frac{T}{T_{\rm bnd}} \spa T_{\rm bnd} = \left( \frac{4\sqrt{3}T_{\rm D3}}{9\pi^2 N} \right)^{\frac{1}{4}}\, .
\end{equation}
We see that \eqref{alphasol} and \eqref{T_delta} provides $\alpha$ in terms of $\overline{T}$, $v$ and $\zeta$.

Consider for a moment the behavior of the solution for $r\rightarrow \infty$. From \eqref{kappa_zeta} we see that $\zeta(r) \rightarrow 0$ for $r\rightarrow \infty$ which physically means the F-string flux goes to zero as the $k$ F-strings are spread out on a larger and larger area.

Define $v_\infty = \lim_{r\rightarrow \infty} v(r)$. If one assumes $0 < v_\infty < 1$ one gets from \eqref{Wdef} that $r_0 \sim 1/r^2$ for $r \rightarrow \infty$. Hence from \eqref{Nquant} one gets $e^\alpha \sim r^4$. But then from \eqref{localtemp} we get $\CT \sim 1/r^2$ which is inconsistent with \eqref{globT}. Thus, we have derived that either $v_\infty=0$ or $v_\infty=1$. The latter case seems consistent, however, it is not the BIon configuration that we aim to study since it has $\CT \rightarrow 0$ for $r\rightarrow \infty$ and therefore presumably asymptotes to an extremal D3-brane at infinity. This seems an interesting alternative scenario to consider. However, we choose here to consider the BIon configuration that asymptotes to a finite-temperature D3-brane for $r\rightarrow \infty$ which hence fixes $v_\infty=0$. 

To summarize, 
\begin{equation}
\zeta (r) \rightarrow 0 \ \ \mbox{for} \ \ r \rightarrow \infty \spa v (r) \rightarrow 0 \ \ \mbox{for} \ \ r \rightarrow \infty\, .
\end{equation}
As a consequence $\CT (r) \rightarrow T$, \ie the local temperature approaches the global temperature far away from the throat of the BIon. This also means that the definition of $T$ is sensible (\ie one could have multiplied it with a constant).
Consider now \eqref{T_delta}. Clearly, the LHS is constant. Instead the RHS for $r\rightarrow \infty$ goes to $\displaystyle \lim_{r \to\infty}\cos \delta$. Thus, we have derived the upper bound on the temperature
\begin{equation}
\overline{T} \leq 1\, .
\end{equation}

We now want to find $v(r)$ from $W$ using \eqref{Wdef}. We compute
\begin{equation}
W = \frac{8}{\pi} \frac{T_{\rm D3}^2}{T_{\rm bnd}^4} \frac{r^2v}{ \cosh^4 \alpha (1-v^2)^3\overline{T}^4} \, .
\end{equation}
Using \eqref{alphasol} and \eqref{T_delta}  we get
\begin{equation}
\label{eq:barW}
\overline{W} = \frac{r^2 v (1-v^2) (1+\frac{\kappa^2}{r^4})}{\sin^2 ( \frac{\delta}{3}+\frac{\pi}{6} )}\, ,
\end{equation}
where we defined the rescaled energy flux
\begin{equation}
\overline{W} = \frac{\sqrt{3}}{2\pi N T_{\rm D3}} \, \frac{W}{\overline{T}^4 }\, .
\end{equation}
We can combine \eqref{eq:barW} with 
\begin{equation}
\label{eq:delta}
\cos \delta = \overline{T}^4 (1-v^2)^2 \sqrt{ 1+\frac{\kappa^2}{r^4} }\, .
\end{equation}
This provides $\overline{W}$ in terms of $v(r)$ and $r$, as well as the constants $\overline{T}$ and $\kappa$. 
Thus, inverting this, we can get the function $v(r)$ in terms of the constants $\overline{W}$, $\overline{T}$ and $\kappa$.
We see now that we have obtained $F(r)$ as written in \eqref{Ffct} in terms of $\overline{W}$, $\overline{T}$ and $\kappa$. 

Notice now that the dependence on $r$ in \eqref{alphasol}, \eqref{eq:barW}, \eqref{eq:delta} and \eqref{Ffct} is only in terms of $r/\sqrt{\kappa}$. Thus, if we regard $F$ as a function of $r/\sqrt{\kappa}$ this function is independent of $\kappa$. Considering \eqref{zofr} we see that $z/\sqrt{\kappa}$ as a function of $r/\sqrt{\kappa}$ does not depend on $\kappa$. Thus, changing $\kappa$ while keeping $\overline{W}$, $\overline{T}$ and $r_{\rm min}/\sqrt{\kappa}$ only changes the solution by a rescaling. Hence, we can set $\kappa=1$ without loss of generality.

We have thus found that the solution \eqref{zofr} is determined completely when given 
\begin{equation}
\overline{T} \spa \overline{W} \spa r_{\rm min}\, ,
\end{equation}
and when setting $\kappa=1$ for simplicity.

\subsection{Velocity distribution along the brane}

We now study the velocity distribution along the radial direction on the brane world-volume. As we already said this can be obtained 
from the energy flux conservation \eqref{eq:barW} along with eq.~\eqref{eq:delta}. Due to the transcendent nature of these equations
we can only work numerically. Without loss of generality we will set $\kappa = 1$ in all the plots and numerical computations.

\begin{figure}[h]
	\centering
	\subfloat[$\overline{T}=0.5$, $\overline{W}=5$]{\includegraphics[scale=.5]{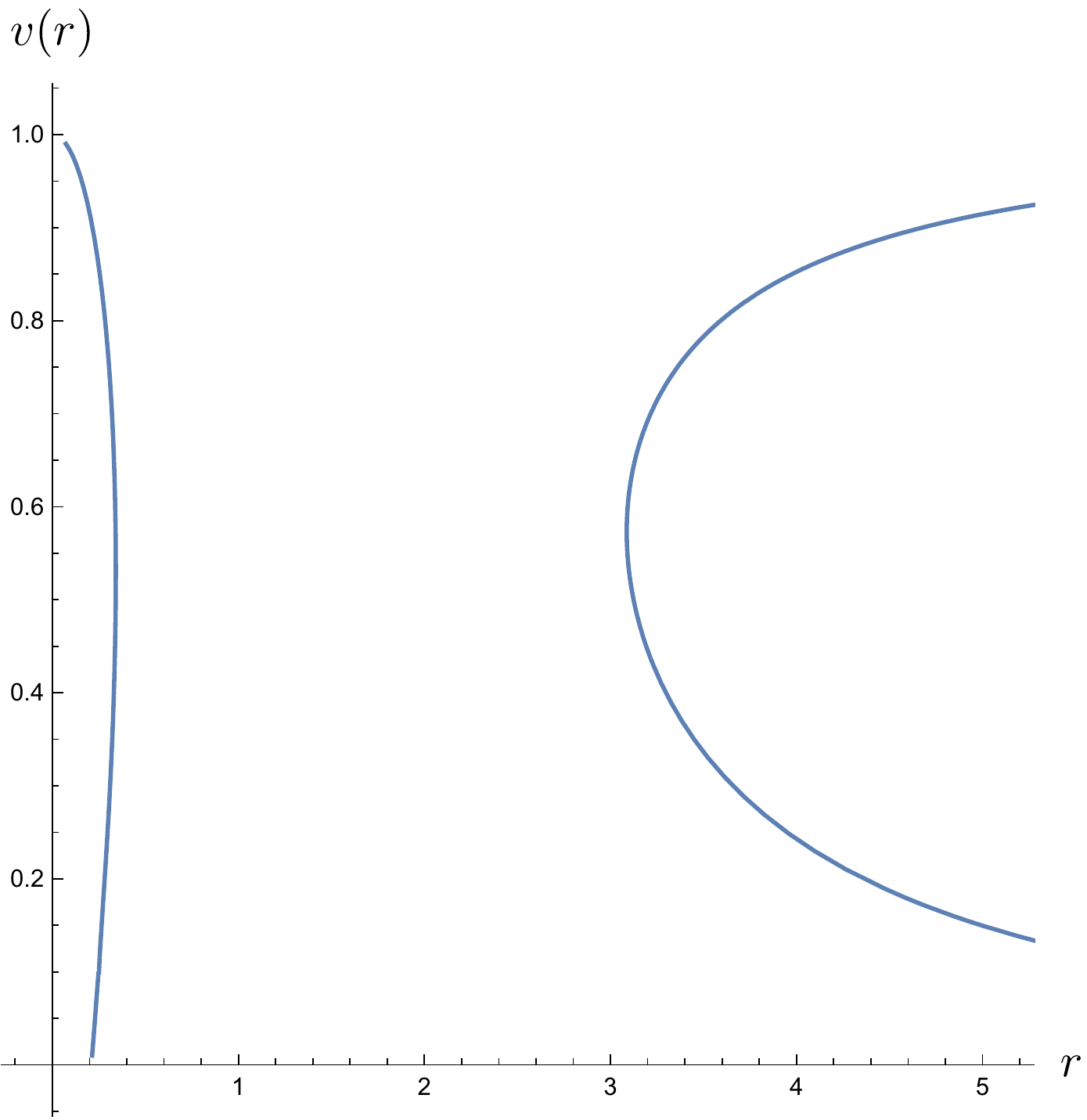}} \quad
	\subfloat[$\overline{T}=0.5$, $\overline{W}=1.6$]{\includegraphics[scale=.5]{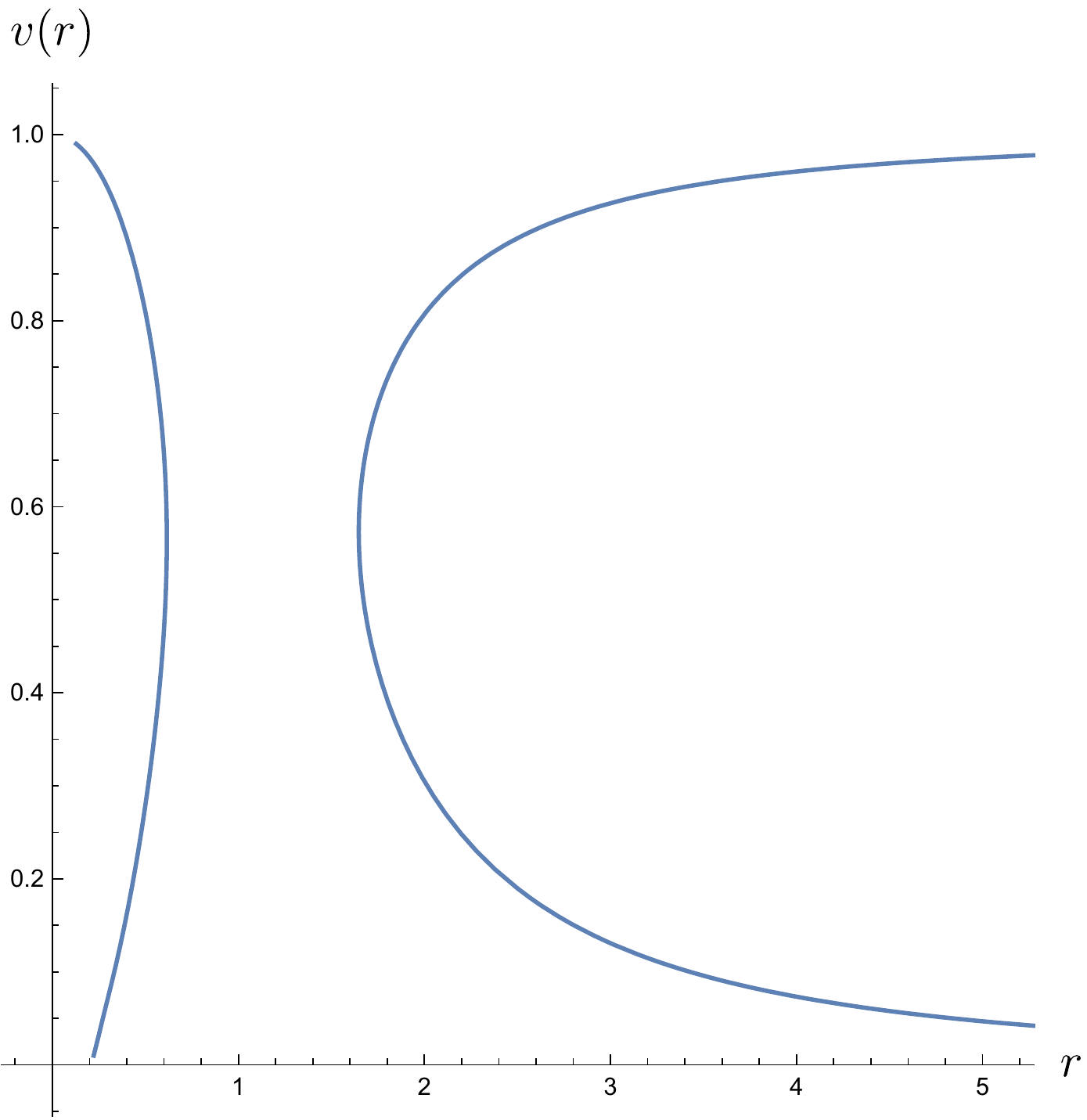}}
	\caption{$v$ as a function of $r$: The plots show the behavior for $\overline{W}>\overline{W}_{\rm crit}$,
		with two branches separated by a gap in the $r$ domain.}
	\label{fig:vvsr}
\end{figure}

Figure~\ref{fig:vvsr} shows two examples of the behavior of the velocity as a function of $r$. This is the behavior that holds 
for large enough $\overline{W}$. We see that we have two disconnected curves separated in the $r$ domain: 
Indeed there is a range of values of $r$ for which the energy flux
constraint cannot be satisfied (for fixed values of $\overline{T}$ and $\overline{W}$). Only the curve on the right is physical, since it is 
the one that reaches $r\to \infty$. The ``turning point'' of this curve, namely the point where $v'(r)$
diverges, is given by the minimal possible value for $r$, which we denote it by $r_{\rm bnd}$.
The curve is formed by two branches that meet at $r_{\rm bnd}$ (for each value of $r>r_{\rm bnd}$ there are two possible values of 
the velocity), but we are interested only in the lower branch since we want to study the case in which $v\to 0$ for $r\to \infty$.
As the two plots suggest the value of  $r_{\rm bnd}$ depends on $\overline{W}$.

\begin{figure}[h]
	\centering
	\subfloat[$\overline{T}=0.5$, $\overline{W}=1.04225 \simeq \overline{W}_{\rm crit}$]{\includegraphics[scale=.5]{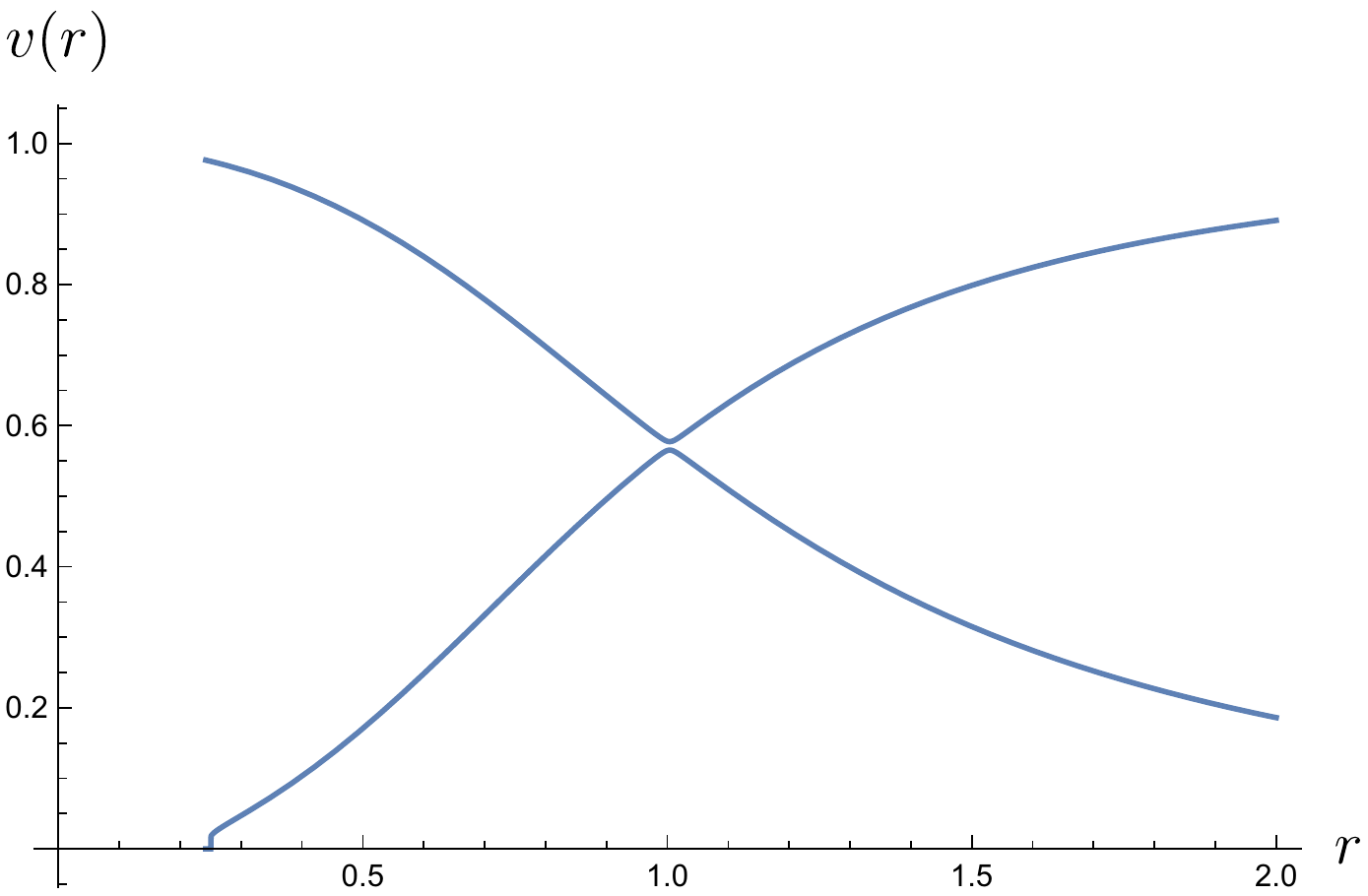}} \quad
	\subfloat[$\overline{T}=0.5$, $\overline{W}=0.8$]{\includegraphics[scale=.5]{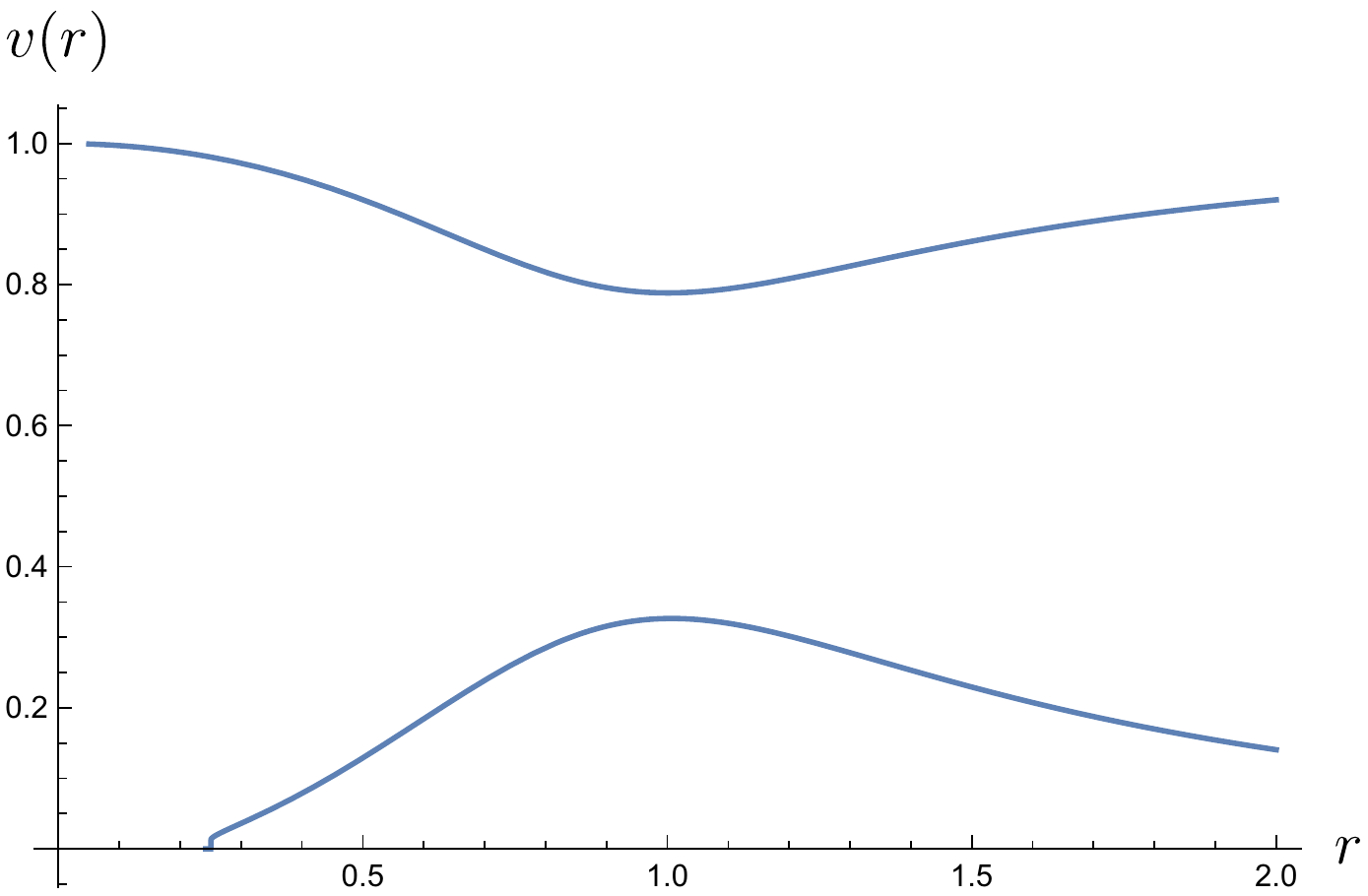}}
	\caption{$v$ as a function of $r$: The plots show the behavior for $\overline{W}\le \overline{W}_{\rm crit}$: We observe that
		also below $\overline{W}_{\rm crit}$ we have two branches, but now separated by a gap in the $v$ domain.}
	\label{fig:vvsr2}
\end{figure}

Decreasing the value of $\overline{W}$ the two curves get closer until for
a particular value of $\overline{W}= \overline{W}_{\rm crit}$ they join. 
For $\overline{W}< \overline{W}_{\rm crit}$ the behavior of the velocity as a function of $r$ changes significantly, and it is shown
in Figure~\ref{fig:vvsr2}. In this regime again we have two disconnected curves but now they are separated by a gap in the velocity domain.
As before the branch we are interested in is the lower one. Also in this case for fixed values of $\overline{T}$ and $\overline{W}$, 
we have a minimum value for $r$, $r_{\rm bnd}$, below which the energy flux cannot be conserved any longer. However, unlike before,
as long as $\overline{W}< \overline{W}_{\rm crit}$, the value of $r_{\rm bnd}$ seems to depend very mildly on $\overline{W}$.

Now we try to understand better the origin of this behavior. Let us rewrite the energy flux conservation equation \eqref{eq:barW} as
\begin{equation}\label{eq:fluxconstr}
v(1-v^2)=\sin^2\left(\frac{\delta}{3}+\frac{\pi}{6}\right) \frac{r^2}{r^4+\kappa^2} \,\overline{W}\, .
\end{equation}
The LHS ranges from 0, which is reached at $v=0$ and $v=1$, to a maximum $\frac{2}{3\sqrt{3}}$ which is reached at $v=\frac{1}{\sqrt{3}}$.
The forbidden region in $r$ in the the plots in Figure~\ref{fig:vvsr} is due to the fact that when $\overline{W}$
is large enough, there are some values of $r$ for which the RHS becomes bigger than $\frac{2}{3\sqrt{3}}$, and the energy flux equation cannot 
be satisfied any longer. 
In order to determine this region, and in particular $r_{\rm bnd}$ we have to study the behavior of the RHS.
This is difficult to analyze analytically because of the term $\sin^2\left(\frac{\delta}{3}+\frac{\pi}{6}\right)$ 
which depends in a highly non trivial way on both $r$ and $v$, through eq.~\eqref{eq:delta}.
Neglecting this term it is straightforward to see that the RHS reaches a maximum at $r=\sqrt{\kappa}$.\footnote{The
	$r$ dependence of the RHS is set by the term $\frac{r^2}{r^4+\kappa^2}$ which goes to 0 for $r \to 0$ and $r\to \infty$
	and reaches a maximum value of $1/(2\sqrt{\kappa})$ at $r=\sqrt{\kappa}$.}
However it turns out that also taking the $\sin^2\left(\frac{\delta}{3}+\frac{\pi}{6}\right)$ 
term into account the behavior in $r$ of the RHS does not get qualitative
modifications. The position of the maximum changes slightly and depends on the temperature but in any case
it always remains close to $r=\sqrt{\kappa}$.
Thus we have that for large enough $\overline{W}$, there is a range of values of $r$ around $r\simeq \sqrt{\kappa}$
for which the energy flux constraint does not admit any solution. Accordingly, the value of $r_{\rm bnd}$ can be determined numerically as the radius $r$ which solves the energy flux constraint for $v=\frac{1}{\sqrt{3}}$, \textit{i.e.} the $r$ which makes the RHS of \eqref{eq:fluxconstr} equal to $\frac{2}{3\sqrt{3}}$.

\begin{figure}[h]
	\centering
	\includegraphics[scale=.7]{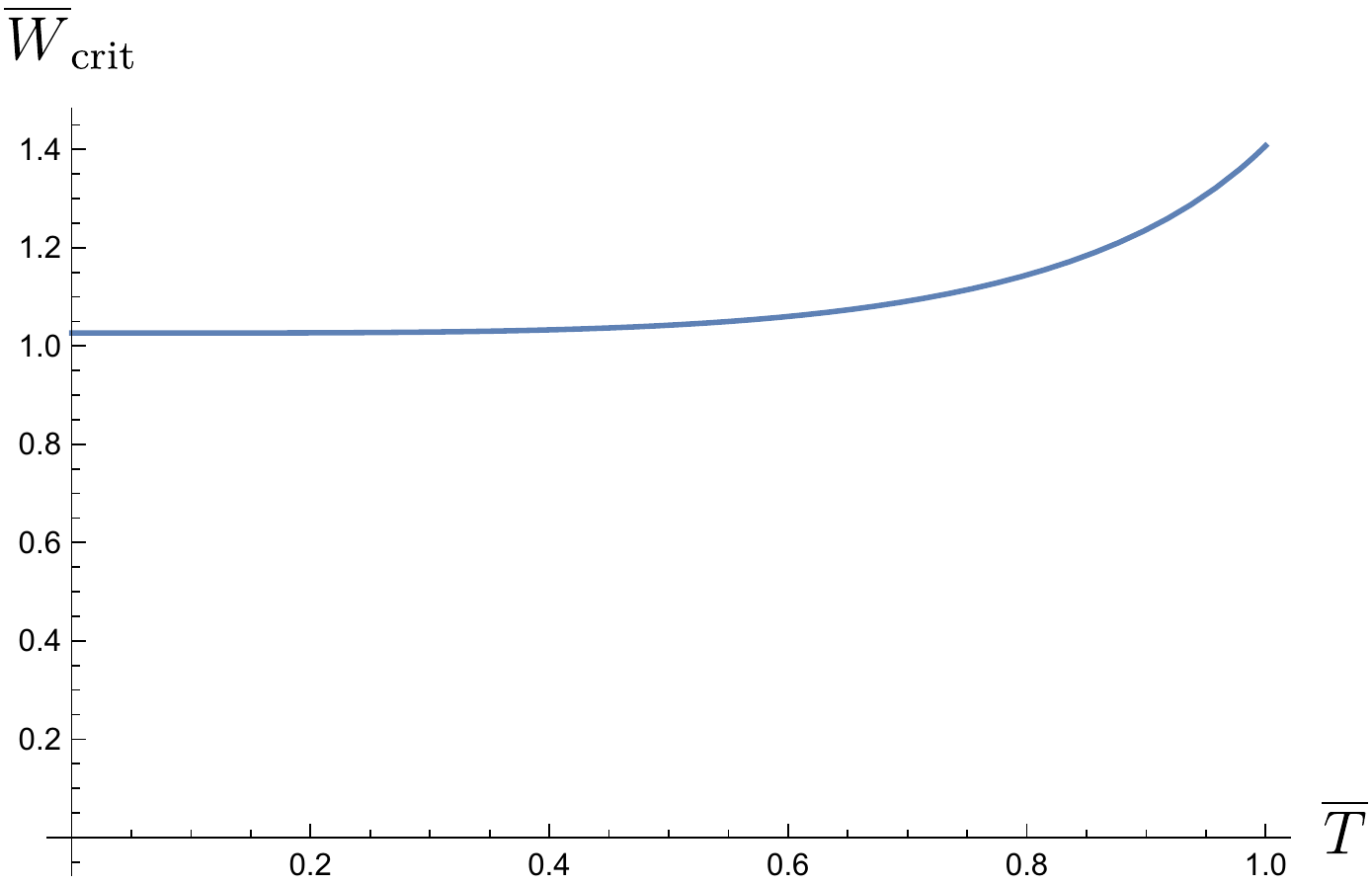}
	\caption{$\overline{W}_{\rm crit}$  as a function of the temperature $\overline{T}$.}
	\label{fig:Wcrit} 
\end{figure}

Following the same reasoning we can also understand what happens when we decrease $\overline{W}$. Since $\overline{W}$ is a multiplicative factor in the 
RHS of \eqref{eq:fluxconstr} when this becomes small enough it makes the whole RHS always smaller than $\frac{2}{3\sqrt{3}}$ and thus the forbidden $r$ region disappear as it is shown in Figure~\ref{fig:vvsr2}. Then the critical value of $\overline{W}$ is the one for which the energy flux conservation constraint \eqref{eq:fluxconstr} evaluated at $v=\frac{1}{\sqrt{3}}$ and at 
the $r$ which makes the RHS maximum, is satisfied.
The behavior of $\overline{W}_{\rm crit}$ as a function of the temperature is shown in Figure~\ref{fig:Wcrit}.

\begin{figure}[h]
	\centering
	\includegraphics[scale=.9]{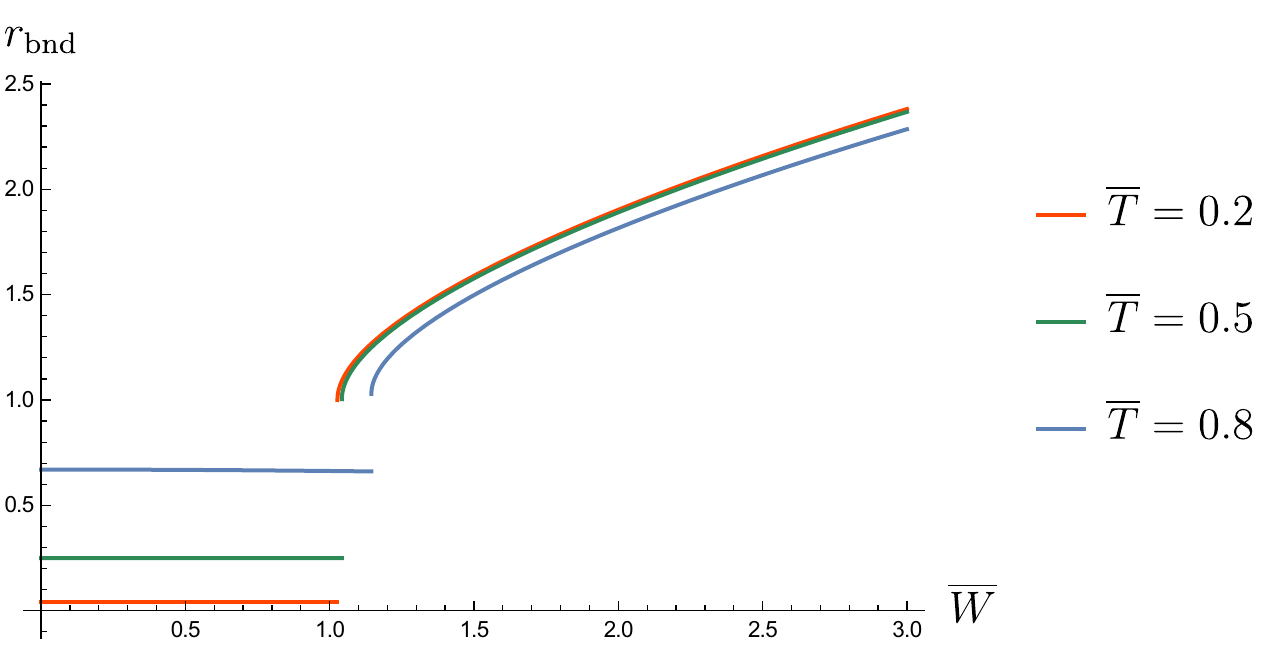}
	\caption{$r_{\rm bnd}$ as a function of $\overline{W}$ for $\overline{T}=0.2$ (orange line),
		$\overline{T}=0.5$ (green line), $\overline{T}=0.8$ (blue line).}
	\label{fig:rbndvsW} 
\end{figure}

Note that for $\overline{W}<\overline{W}_{\rm crit}$ we still have a lower bound on $r$ which in this regime follows from
the fact that the expression giving $\cos \delta$  in eq.~\eqref{eq:delta} is bounded from above to 1,
\begin{equation}
\overline{T}^4 \left(1-v^2\right)^2\sqrt{1+\frac{\kappa^2}{r^4}}\le 1\, .
\end{equation}
This is equivalent to the condition $r\ge r^*$ with
\begin{equation}
\label{rstar}
r^*= \sqrt{\kappa} \, \overline{T}^2\left(1-v^2\right)\left(1-\left(1-v^2\right)^4\overline{T}^8\right)^{-\frac{1}{4}}\, .
\end{equation}
Of course this bound on $r$ is always valid also when $\overline{W}>\overline{W}_{\rm crit}$, but in this case it 
turns out that $r_{\rm bnd}$ defined above is always larger that $r^*$. We thus have that $r_{\rm bnd}$, the minimum possible value for
$r$, is given overall by
\[
r_{\rm bnd} = \left\{ 
\begin{array}{lr}
\mbox{value of $r$ that satisfies \eqref{eq:fluxconstr} with } v=\frac{1}{\sqrt{3}}, &  \overline{W}>\overline{W}_{\rm crit} \\
\mbox{value of $v$ that satisfies \eqref{eq:fluxconstr} with $r=r^*$}, \, &  \overline{W}\le\overline{W}_{\rm crit}
\end{array}
\right.
\]
The behavior of $r_{\rm bnd}$ as a function of $\overline{W}$ is shown in Figure~\ref{fig:rbndvsW}. We can see that at $\overline{W}_{\rm crit}$ 
the value of $r_{\rm bnd}$ jumps discontinuously and this signals that the system undergoes a phase transition.
%


\subsection{Brane separation}

\begin{figure}[h]
	\centering
	\subfloat[$\overline{W}=0.8 $\label{fig:DeltaA}]{\includegraphics[scale=.75]{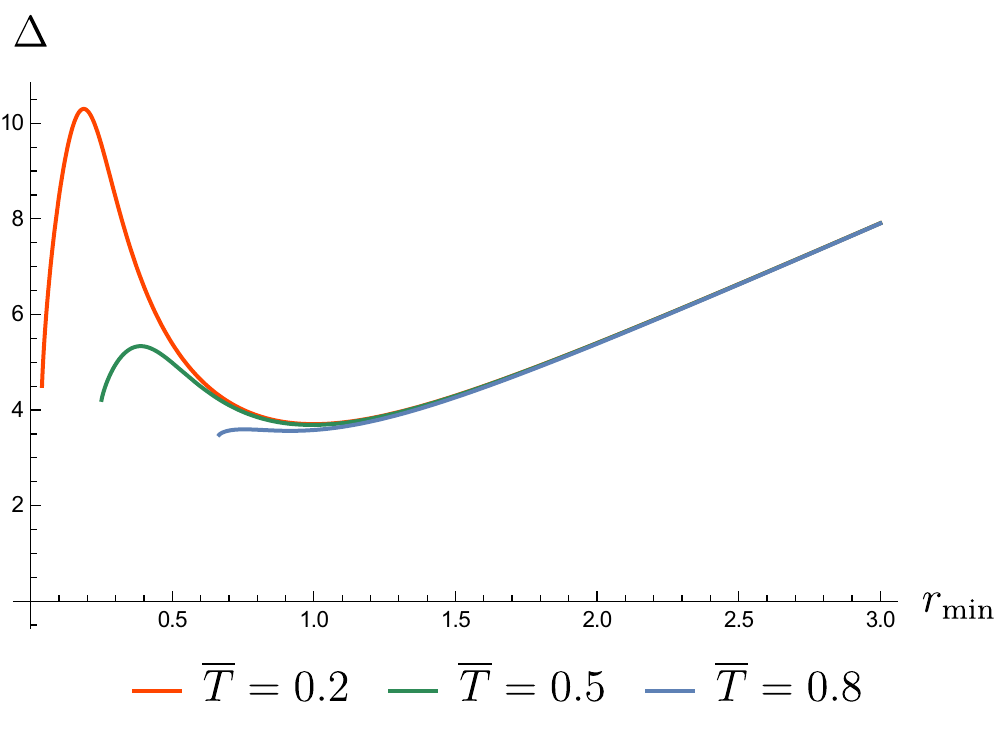}} \;
		\subfloat[$\overline{W}=1.6 $]{\includegraphics[scale=.75]{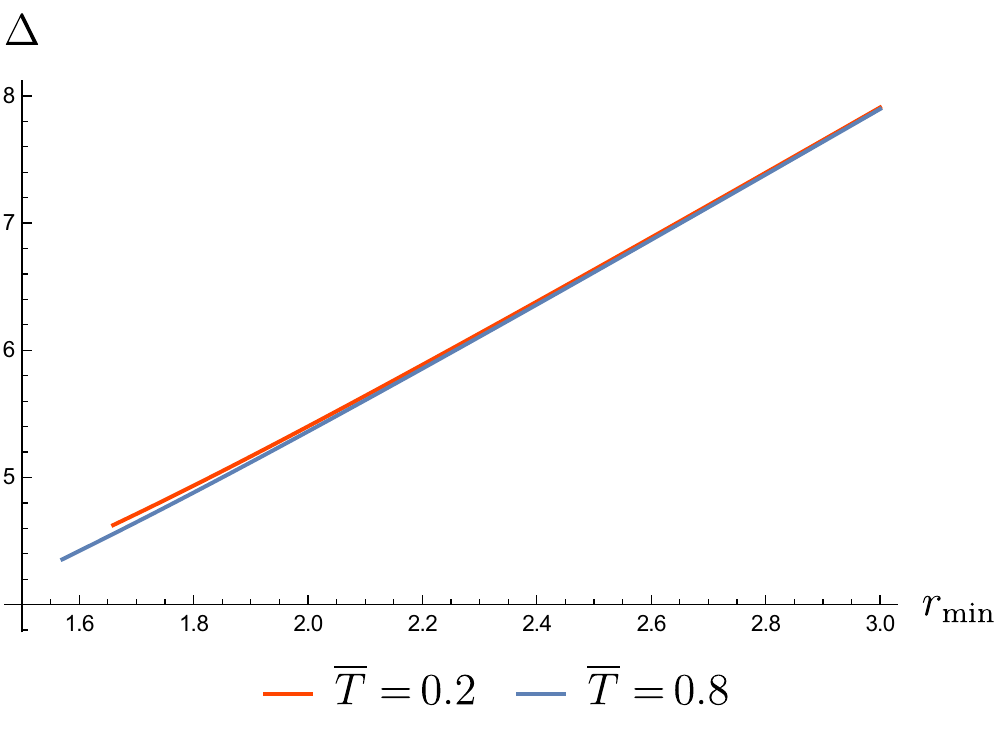}}
		\caption{$\Delta$ as a function of $r_{\rm min}$.}
		\label{fig:Delta}
	\end{figure}

We saw that the solution of the extrinsic equation \eqref{zofr}
is written in terms of a parameter $r_{\rm min}$ which gives the minimum radius of the throat in the BIon configuration.
By definition this is the point where the derivative of the profile function, $z'(r)$ diverges.
$r=r_{\rm min}$ defines the surface where we have to glue the brane and the anti-brane world-volumes together. In order to
have a smooth configuration the anti-brane solution is built as the mirror of the brane solution. 
The separation between the branes is then given by \cite{Grignani:2010xm}
\begin{equation}
\Delta=2\int_{r_{\rm min}}^\infty dr' \left(\frac{F(r')^2}{F(r_{\rm min})^2}-1\right)\, ,
\end{equation}
and it is plotted as a function of $r_{\rm min}$ in Figure~\ref{fig:Delta} for different values of $\overline{T}$. Note that above $\overline{W}_{\rm crit}$ there is only one 
phase for each (allowed) value of $\Delta$, while below $\overline{W}_{\rm crit}$ we can distinguish two cases: 
For temperatures $\overline{T}\le \overline{T}_{b}$ with $\overline{T}_b \simeq 0.8$ there are
three branches in the plot $\Delta$-vs-$r_{\rm min}$. As shown by the orange line in Figure~\ref{fig:DeltaA}, 
in this regime there appear a local maximum and a local minimum for $\Delta$. 
Let us denote them as $\Delta_{\rm max}$ and $\Delta_{\rm min}$ respectively. For a fixed value of $\Delta$ in the
range $\left(\Delta_{\rm min},\Delta_{\rm max}\right)$ there corresponds three different values of $r_{\rm min}$, \ie 
three different BIon solutions. For temperatures above $\overline{T}_b$, $\Delta_{\rm max}$ and $\Delta_{\rm min}$ cease
to exist and, just like for $\overline{W}>\overline{W}_{\rm crit}$, we have only one possible phase for each value of $\Delta$. 
Basically, for $\overline{W}<\overline{W}_{\rm crit}$ the behavior seems to be completely analogous to the one that holds in
the zero velocity ($\overline{W}=0$) case \citep{Grignani:2010xm,Grignani:2011mr}. 
	%

\section{Action principle and thermodynamics}
\label{sec:action}

In this section we generalize the thermodynamics of the static thermal BIon solution to the case of the flowing BIon. In particular, we introduce a new thermodynamic quantity called the integrated velocity as well as a total entropy. The thermodynamic quantities obey a first law, and they are connected to the free energy from which we can get an action principle for the BIon. In addition, we regularize the free energy and consider the free energy of the phases of the flowing BIon.

\subsection{Action principle}

We want to generalize the thermodynamic action, \ie  the free energy, of the hot BIon solution with $W=0$ of Ref.~\cite{Grignani:2010xm} to non-zero $W$. In the $W=0$ case the free energy is
\begin{equation}
\label{CF_zeroW}
\CF = \int dV_{(3)} (\epsilon - \CT s) = - \int dV_{(3)} P_\parallel\, ,
\end{equation}
where
\begin{equation}
\int dV_{(3)} g(r) = 4\pi \int_{r_{\rm min}}^\infty dr\, r^2 \sqrt{1+{z'}^2 } g(r) \, .
\end{equation}
We want to find a generalized expression for the free energy that should reduce to the above for $W=0$ and that should give the extrinsic equation \eqref{extreq} for any $W$. We propose
\begin{equation}
\label{CF_gen}
\CF = - \int dV_{(3)} \gamma_{11} T^{11} = - \int dV_{(3)} \Big( P_\parallel + \frac{v^2}{1-v^2} \CT s \Big)\, .
\end{equation}
We note first that this reduces to \eqref{CF_zeroW} for $W=0$ since $v(r)=0$ in that case. To obtain the EOMs for $\CF$ seen as an action we should vary $\CF$ holding $T$ and $W$ fixed (as well as $k$ and $N$). 
We have shown in Sec.~\ref{sec:blackfoldeqs} that $\gamma_{11}T^{11}$ as function of $r$ is uniquely determined given $T$, $W$, $k$ and $N$. This does not change when varying $\CF$.
The only thing that changes when varying $\CF$ is the $\sqrt{1+{z'}^2}$ factor in the volume element $dV_{(3)}$.
Writing the Lagrangian as
\begin{equation}
\CL = 4 \pi r^2 \sqrt{1+{z'}^2 } \gamma_{11} T^{11}\, ,
\end{equation}
the Lagrange equation is
\begin{equation}
\left( \frac{\partial \CL}{\partial z'} \right)' = \frac{\partial \CL}{\partial z}\, ,
\end{equation}
which indeed gives the extrinsic equation \eqref{extreq}. Finally, we notice that the free energy \eqref{CF_gen} generalizes the free energy in a more physical sense, namely that in both cases they are minus the integral over the pressure in the $r=\sigma^1$-direction (on the world-volume). Thus, it seems a very natural generalization from a physical point of view.

\subsection{Thermodynamics}

We aim here to find the thermodynamics that should correspond to regarding \eqref{CF_gen} as a thermodynamical free energy. In the $W=0$ case (for fixed $k$ and $N$) the free energy \eqref{CF_zeroW} satisfies the first law of thermodynamics $d \CF = - S dT$ \cite{Grignani:2011mr}. This relation follows from the local relation $dP_\parallel=s d\CT$. To find the thermodynamics for non-zero $W$ we should add a term corresponding to the variation of $W$, \ie  $d\CF = - S dT - V dW$ where $V$ is the conjugate variable to $W$. As seen from equation \eqref{Wdef} $W$ is the total energy flux passing through a given surface of constant $r$. Thus, $W$ is an extensive variable with respect to the two-sphere directions (transverse to the F-string) and an intensive variable with respect to the radial direction $r=\sigma^1$ (parallel to the F-string). The conjugate variable $V$ should thus be intensive with respect to the two-sphere directions and extensive with respect to the radial direction. Thus, we can write it as
$$V = \int_{r_{\rm min}}^\infty dr \sqrt{1+{z'}^2} \, x(r)$$
where $x(r)$ is an unknown function.
Using this we can write a local version of $d\CF = - S dT - V dW$ as
\begin{equation}
- d \Big( P_\parallel + \frac{v^2}{1-v^2} \CT s \Big) = - \sqrt{1-v^2} \, s \, d \Big( \frac{\CT}{\sqrt{1-v^2}} \Big) - x \, d \Big( \frac{v}{1-v^2} \CT s \Big)\, .
\end{equation}
Using that  $dP_\parallel = s d\CT$ we can rewrite the previous equation as
%
\begin{equation}
- \CT s \, d \Big( \frac{v^2}{1-v^2} \Big)  - \frac{v^2}{1-v^2} d ( \CT s) = - \CT s \sqrt{1-v^2} \, d \Big( \frac{1}{\sqrt{1-v^2}} \Big) - \CT s x \, d \Big( \frac{v}{1-v^2} \Big) - \frac{x v}{1-v^2} d( \CT s) \, .
\end{equation}
We see now that if we set $x=v$ then the $d( \CT s)$ terms cancel with each other and we are left with the following relation
\begin{equation}
 d \Big( \frac{v^2}{1-v^2} \Big)  =  \sqrt{1-v^2} \, d \Big( \frac{1}{\sqrt{1-v^2}} \Big) +v \, d \Big( \frac{v}{1-v^2} \Big) \, ,
\end{equation}
which is identically satisfied. Thus, we have shown the identity
\begin{equation}
- d \Big( P_\parallel + \frac{v^2}{1-v^2} \CT s \Big) = - \sqrt{1-v^2} \, s \, d \Big( \frac{\CT}{\sqrt{1-v^2}} \Big) - v \, d \Big( \frac{v}{1-v^2} \CT s \Big)\, .
\end{equation}
Integrating this expression gives us the first law of thermodynamics for the free energy $\CF$ defined in \eqref{CF_gen}, namely
\begin{equation}
\label{firstlawCF}
d\CF = - S dT - V dW \ \ \ \ \mbox{(for fixed $k$ and $N$)}\, ,
\end{equation}
where $V$, which we call the integrated velocity, is given by
\begin{equation}
\label{Vdef}
V = \int_{r_{\rm min}}^\infty dr \sqrt{1+{z'}^2} \, v(r)\, ,
\end{equation}
and the total entropy $S$ is given by
\begin{equation}
\label{genS}
S = \int dV_{(3)} \sqrt{1-v^2} s\, .
\end{equation}
This shows that $\CF=\CF(T,W,k,N)$ is a true thermodynamic potential. Note that since we already had identified $\CF$, $T$ and $W$ beforehand, the derived definitions of $S$ and $V$ follows uniquely from \eqref{firstlawCF}. Thus, even if it might seem like we made guesses on the way to finding $S$ and $V$, the result is unique.

We can now find the mass/energy potential $M=M(S,V,k,N)$ that depends on $V$ which is extensive in the radial direction. To get this we perform the Legendre transform $M = \CF + TS + VW$ giving%
\footnote{We note that this is not the only quantity that generalizes the mass for $W=0$. Also the potential $E = \CF + TS = M - VW$ is a generalization,
\begin{equation}
E = \int dV_{(3)} \Big( \epsilon - \frac{v^2}{1-v^2} \CT s \Big)\, ,
\end{equation}
with $dE = T \, dS - V \,dW$ for fixed $k$ and $N$. However, for our purposes the sphere directions do not play an important role as our configuration is spherically symmetric. Hence it is more natural to consider the mass to be extrinsic with respect to the radial direction.}
\begin{equation}
M = \int dV_{(3)} \epsilon \, ,
\end{equation}
\begin{equation}
\label{firstlaw}
dM = T \, dS + W \,dV \ \ \ \ \mbox{(for fixed $k$ and $N$)}\, .
\end{equation}
With this we can perform a consistency check on the thermodynamics. The action is $I = \beta \CF = \beta ( M-TS-VW)$  \cite{Grignani:2011mr}. From the above we know that the action principle $\delta I / \delta z(r)=0$ gives the extrinsic equation \eqref{extreq} under variations with fixed $T$ and $W$. Hence, we can infer that it must also follow from
\begin{equation}
\label{firstlawvary}
\frac{\delta M}{\delta z(r) } - T \frac{\delta S}{\delta z(r)} - W \frac{\delta V}{\delta z(r)} = 0\, ,
\end{equation}
which is a variational form of the first law of thermodynamics \eqref{firstlaw} (even if we derived \eqref{firstlaw} using only the local thermodynamic variations keeping $z(r)$ fixed).
We compute
\begin{equation}
\begin{array}{c}
\frac{\delta M}{\delta z(r) } = - \left( \frac{z'}{\sqrt{1+{z'}^2}} 4\pi r^2 \epsilon \right)' \spa \frac{\delta S}{\delta z(r) } = - \left( \frac{z'}{\sqrt{1+{z'}^2}} 4\pi r^2  \sqrt{1-v^2}s \right)'
\spa
\frac{\delta v}{\delta z(r) } = - \left( \frac{z'}{\sqrt{1+{z'}^2}}  v \right)'\, .
\end{array}
\end{equation}
Hence from Eq.~\eqref{firstlawvary} we obtain
\begin{equation}
\left( \frac{z'}{\sqrt{1+{z'}^2}} \Big[ 4\pi r^2 \epsilon - T 4\pi r^2 \sqrt{1-v^2} s - W v\Big] \right)' =0\, ,
\end{equation}
which indeed is equivalent to \eqref{extreq}.


\subsection{Comparison of free energies}

We are now interested in determining which phase is thermodynamically favored. In order to achieve this 
we have to compare the free energy of the solutions. We work in an ensemble in which the quantities $(T,W,\Delta,k,N)$
are held fixed. In this case, as we argued in the previous section, the free energy $\CF$ defined in \eqref{CF_gen} is the 
suitable thermodynamic potential we have to use. Actually, in all the numerical computations of free energy
we will set $N=1$ and $\kappa =1$, thus we will study the thermodynamic stability of the solution
only as a function of the temperature $T$, of the energy flux $W$ and of the separation $\Delta$.

In order to explicitly compute the free energy of a particular configuration of the brane/anti-brane system it is convenient 
to rewrite $\CF$ given in Eq.~\eqref{CF_gen} as
\begin{equation}
	\CF= 2 W \int_{r_{\rm min}}^{\infty} dr \frac{F(r)^2}{\sqrt{F(r)^2-F(r_{\rm min})^2}}\, ,
\end{equation}
and then evaluate the latter on the solution.
This is clearly divergent because of the infinite extension of the branes and it has to be regularized.
Along the lines of \cite{Grignani:2011mr}, we
introduce a fictitious parameter $r_{\rm cut}$ and define a regularized free energy $\delta\CF$ by 
subtracting to the free energy of any solution that of the solution with $r_{\rm min}=r_{\rm cut}$
\begin{equation}
	\delta \CF[r_{\rm min}] = \CF[r_{\rm min}] - \CF[r_{\rm cut}]\, .
\end{equation}

\begin{figure}[h]
	\centering
	\subfloat[$\overline{W}=0.8$, $r_{\rm cut}=2.5$]{\includegraphics[scale=.8]{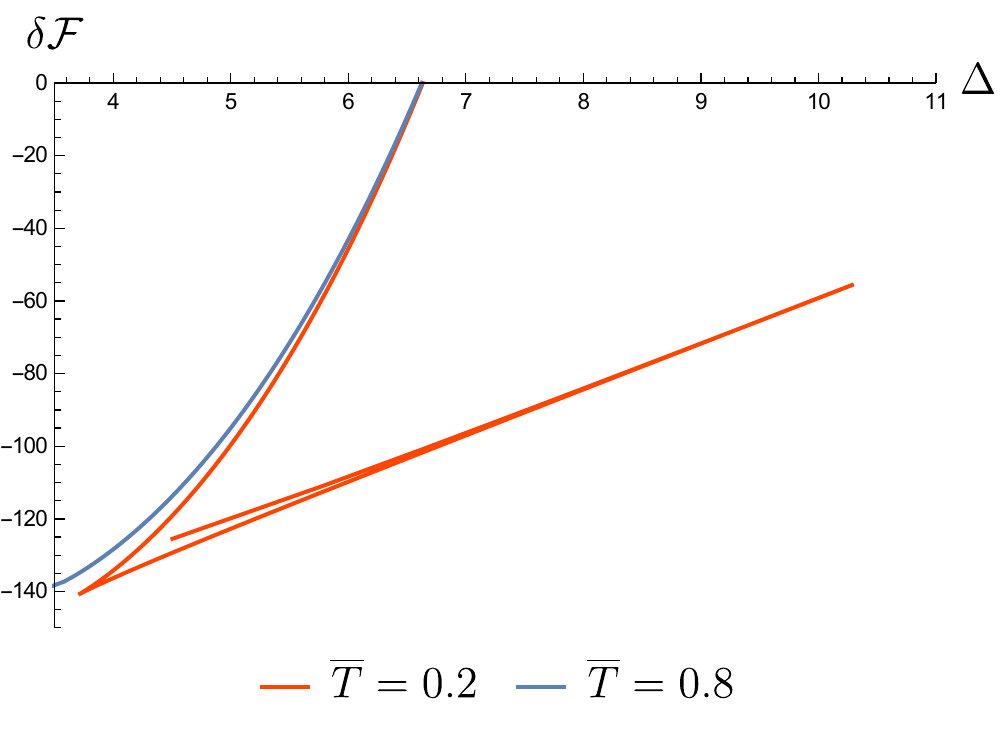}}
	\subfloat[$\overline{W}=1.6$, $r_{\rm cut}=2.5$]{\includegraphics[scale=.8]{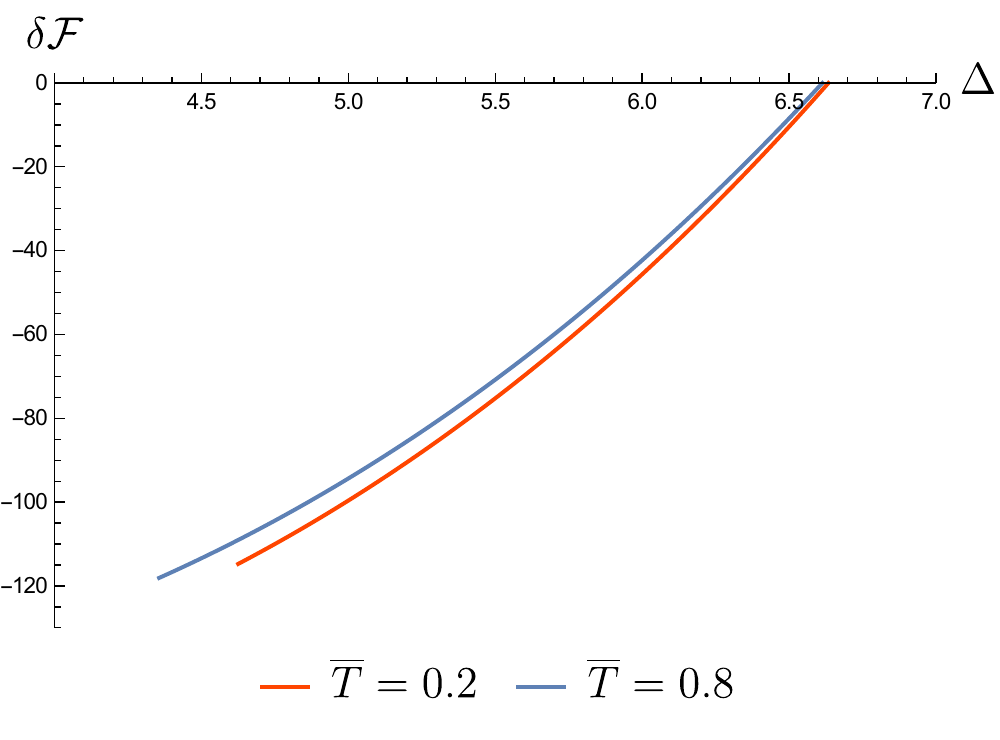}}
	\caption{Regularized free energy $\delta\CF$ as a function of $\Delta$.}
	\label{fig:Free_energy}
\end{figure}

Figure~\ref{fig:Free_energy} shows the behavior of the free energy $\delta\CF$ as a function of the brane/anti-brane 
separation $\Delta$ for two values of the temperature and two values of $\overline{W}$, one below and one above $\overline{W}_{\rm crit}$ ($\overline{W}=0.8$  and $\overline{W}=1.6$ respectively).
Again we see that below $\overline{W}_{\rm crit}$ the behavior is similar to the case without flux \cite{Grignani:2011mr}.
%


\section{Conclusions}
\label{sec:conclusion}

In this paper we have studied the thermal BIon configuration of \cite{Grignani:2010xm} in the presence of a radial boost along the brane,  using the blackfold approach.
We found that the presence of the boost implies that the thermal BIon acquires a non-zero energy flow $W$ which is conserved along the radial direction. 

A very interesting feature of the thermal BIon with heat flow, which we dubbed ``flowing BIon", is that with the presence of the heat flow the configuration is a non-stationary blackfold system. This is a new example of a black brane configuration with an event horizon that is not a Killing horizon \cite{Caldarelli:2008mv,Emparan:2009at}. This opens up the possibility to find other new black brane configurations in higher dimensions that do not have Killing horizons using the blackfold formalism.

In this paper we initiated this type of application of the blackfold formalism by looking at the lowest order in the blackfold perturbative approach. To this order our description is not able to capture dissipative effects, therefore it would be very interesting to go further with the perturbative approach and thus include the effect of dissipation in the BIon configuration. This could possibly mean that the two heat reservoirs in the flowing BIon configuration are no longer at the same temperature.

By studying of the flowing BIon system we have discovered a new global thermodynamics finding the free energy for a system with integrated velocity $V$ and heat flow $W$. This is done using the blackfold description. We also found that the flowing BIon undergoes a phase transition when the energy flux $W$ reaches a critical value $W_{\rm crit}$. While the system behaves qualitatively similar to the non-flowing solution when $W<W_{\rm crit}$, for $W>W_{\rm crit}$ we instead notice a drastic change with respect to the $W=0$ case in that the separation between the branes in the system grows exponentially as function of $W$.

More generally, we have possibly found a new type of blackfold configuration with heat flow that could be interesting to study in generality and for other cases. It is conceivable that one could develop a general framework for blackfolds with heat flow, possibly showing that the relation between the global temperature and the local temperature that we found for the case of the BIon is a general feature.


 \section*{Acknowledgments}

We thank Jay Armas and Jorge Santos for useful discussions.
T.H.\ acknowledges support from FNU grant number DFF-6108-00340 and the Marie-Curie-CIG grant number 618284 and thanks Perugia University for their great hospitality.

\begin{appendix}

\section{More on relativistic fluid temperature}
\label{app:fluidtemp}

We consider the most simple setting possible: A perfect fluid in Minkowski space-time. Hence the energy-momentum tensor is
\begin{equation}
T^{ab} = (\epsilon + P ) u^a u^b + P \eta^{ab} = \CT s u^a u^b + P\eta^{ab}
\end{equation}
where $u^a(x)$ is the fluid velocity (with $u^2 = -1$) as observed by a global observer who also observes the energy-momentum tensor $T^{ab}$. We consider a stationary situation, thus $\partial_t u^a = 0$, $\partial_t \epsilon= 0$ and $\partial_t P=0$. In the second equation we used the Gibbs-Duhem relation $\epsilon + P = \CT s$. Here $\CT$ is the temperature measured by an observer that is comoving with the fluid, \ie  in the local rest frame of the fluid. $s$ is the entropy density measured in the local rest frame. 

Conservation of energy-momentum is $\partial_a T^{ab} = 0$. The zeroth component is
\begin{equation}
\label{comp0}
\partial_a T^{a0} = \partial_a ( \CT s u^a u^0 + P \eta^{a0} ) = \partial_a ( \CT s u^a u^0 )
\end{equation}
The component from projecting on $u_b$ is
\begin{eqnarray}
\label{compu}
u_b \partial_a T^{ab} &=& \partial_a ( - \CT s u^a + P u^a ) - T^{ab} \partial_a u_b \nn \\ &=& - \CT  \partial_a ( s u^a) + u^a ( - s \partial_a \CT + \partial_a P  ) - \CT s u^a u^b \partial_a u_b \nn \\ &=& - \CT  \partial_a ( s u^a)
\end{eqnarray}
giving the conservation of the entropy current.
Hence combining \eqref{comp0} and \eqref{compu} we get
\begin{equation}
\label{eqsit1}
u^a \partial_a ( \CT u^0 ) = 0
\end{equation}
Thus, $\CT u^0$ is constant along the direction of the fluid velocity. If one has a situation in which $u^0$ varies along the fluid flow then it makes sense to define
\begin{equation}
\label{Tsit1}
T = \CT u^0
\end{equation}
since this then is the constant temperature in the system.

\subsubsection*{Situation with a Killing vector}

Consider having a fluid configuration for which the fluid velocity is along a time-like Killing vector field $k^a$
\begin{equation}
u^a = \frac{k^a}{|k|}
\end{equation}
One can then derive in a blackfold setting
\begin{equation}
T = |k| \CT
\end{equation}
For blackfolds that are associated to a black holes with a KVF one has (for instance)
\begin{equation}
k = \partial_t + v \partial_x
\end{equation}
Hence 
\begin{equation}
\label{Tsit2}
T = \sqrt{1-v^2} \CT
\end{equation}

\subsubsection*{Comparison}

The two situations are mutually exclusive. The situation for \eqref{Tsit1} requires that there is a non-trivial profile for the solution along the direction of the fluid flow. Otherwise \eqref{eqsit1} does not contain any constraint on the system. But having $u^a \propto k^a$ as in the situation for \eqref{Tsit2} precisely means that all observables are invariant along the fluid flow.

\end{appendix}

\providecommand{\href}[2]{#2}\begingroup\raggedright\endgroup




\end{document}